\documentclass[a4paper,12pt]{article}
\pdfoutput=1
\usepackage{amssymb}
\usepackage{epsfig}
\usepackage{graphicx}
\usepackage{slashed}

\makeatletter

\@addtoreset{equation}{section}
\makeatother
\def\cycl{{\mbox{cycl}}}
\def\be{\begin{equation}}
\def\ee{\end{equation}}
\def\bea{\begin{eqnarray}}
\def\eea{\end{eqnarray}}

\def\({\left(}
\def\){\right)}
\def\<{\left<}
\def\>{\right>}

\def\tr{{\mbox{tr}}}
\def\be{\begin{equation}}
\def\ee{\end{equation}}
\def\bea{\begin{eqnarray*}}
\def\eea{\end{eqnarray*}}
\def\ben{\begin{eqnarray}}
\def\een{\end{eqnarray}}
\def\({\left(}
\def\){\right)}
\def\<{\left<}
\def\>{\right>}
\def\!{\right|}
\def\|{\left|}

\def\[{\left[}
\def\]{\right]}

\def\+{\bar}
\def\mb{\mathbb}
\def\tr{{\mbox{tr}}}

\def\t{\widetilde}

\def\O{{\cal{O}}}

\def\Q{{\cal{Q}}}

\def\l{{{\ell}}}

\begin{document}

\setlength{\unitlength}{1mm}

\pagestyle{empty}
\vskip-10pt
\vskip-10pt
\hfill 
\begin{center}
\vskip 3truecm
{\Large \bf
Superconformal indices on $S^1\times (S^5/\mb{Z}_p)$}
\vskip 2truecm
{\large \bf
Andreas Gustavsson}
\vspace{1cm} 
\begin{center} 
Department of Physics and Astronomy, Uppsala University,\\
Box 516, SE-75120 Uppsala, Sweden
\end{center}
\vskip 0.7truecm
\begin{center}
(\tt agbrev@gmail.com)
\end{center}
\end{center}
\vskip 2truecm
{\abstract{We obtain generating functions associated to the abelian superconformal indices for 6d $(1,0)$ tensor and hypermultiplets on $S^1\times (S^5/\mb{Z}_p)$. We extract the superconformal indices and their high and low temperature behaviors. We consider round and generically squashed $S^5$ in turn. We show that the unsquashed limit of the superconformal indices is smooth. We examine S-duality in the large $p$ limit that acts by exchanging the Hopf circle with the temporal circle.}}

\vfill
\vskip4pt
\eject
\pagestyle{plain}

\section{Introduction}
The partition function for 5d SYM was first computed perturbatively (i.e. by suppressing instanton contributions) on a round $S^5$ in \cite{Kallen:2012va}, \cite{Kim:2012ava}. Later this result was extended to the instanton sector in \cite{Lockhart:2012vp}, \cite{Kim:2012qf} by regularizing with generic squashing parameters $a,b,c$ subject to the relation $a+b+c=0$. For abelian gauge group, the 5d partition function including the instanton sector, was matched with the 6d superconformal index on $S^1\times S^5$ in \cite{Kim:2012qf}, providing supporting evidence for the M5/D4 correspondence \cite{Douglas:2010iu}, \cite{Lambert:2010iw}. 

It remains difficult to generalize the abelian tests of the M5/D4 correspondence to nonabelian gauge groups (except for the case of infinite rank gauge group where we may use the AdS/CFT correspondence) since that requires a definition of the nonabelian M5 brane theory itself.\footnote{However, some recent tests have been made in this direction. The half-BPS index has been computed for $A_k$ 5d SYM by localization in \cite{Kim:2013nva} and corresponding expressions for the $D_k$ and $E_k$ gauge groups have been conjectured in \cite{Beem:2014kka}. These half-BPS indices have been recently deconstructed for the $A_k$ and the $D_k$ series from corresponding 4d quiver theories in \cite{Hayling:2017cva}, \cite{Bourget:2017sxr}. I thank A.~Bourget for bringing these references to my attention.} But we can also try to generalize the test to other geometries while keeping the gauge group abelian. As a first step in that direction, in this paper we will generalize the computation of the abelian 6d superconformal index to $S^1\times (S^5/\mb{Z}_p)$. Keeping $(2,0)$ supersymmetry of the M5 brane leads to a very restricted class of possible 6d geometries. But these lens space geometries belong to that class. The lens spaces have non-trivial topology that is detected by the Ray-Singer torsion. In \cite{Bak:2017jwd} we found a mismatch (a factor that is related to the Ray-Singer torsion) between the 6d index and the 5d partition function, for the maximally topologically twisted 5d maximally supersymmetric Yang-Mills. This mismatch was traced to the nontrivial circle reduction of the selfdual two-form from 6d to 5d. It is then natural to ask whether one can find the same type of mismatch also between the 6d $(1,0)$ theories on $\mb{R} \times (S^5/{\mb{Z}_p})$ and 5d SYM on $S^5/{\mb{Z}_p}$.

The abelian 6d superconformal index on $S^1\times (S^5/\mb{Z}_p)$ has been computed previously in \cite{Kim:2013nva}. We reproduce that result in eq (\ref{Kimyeong}). We then use this result to extract its high temperature expansion, which is what we should get when we compute the partition function for 5d SYM theory on $S^5/{\mb{Z}_p}$. On the other hand, in \cite{Kim:2013nva} the low temperature expansion of this index was found to match with the partition function of 5d SYM theory on $\mb{R}\times \mb{C}P^2$. 

The abelian M5 brane superconformal index on $S^1 \times S^5$ was first obtained in \cite{Bhattacharya:2008zy} by using radial quantization. We reproduced and generalized this in \cite{Bak:2016vpi} by using Hamiltonian quantization. Two types of squashings were considered, squashing of the fiber and of the base manifold respectively. These squashings were also independently found in \cite{Benvenuti:2016dcs}. Three squashing parameters $A,B,C$ are introduced associated to the three Cartan rotations of $SO(6)$, which is the isometry group of $S^5$. The trace parameter 
\bea
h &=& \frac{1}{3}(A+B+C)
\eea
squashes the Hopf fiber of $S^5$. The three traceless squashing parameters 
\bea
a &=& A - h\cr
b &=& B- h\cr
c &=& C-h
\eea
which are subject to the constraint $a+b+c=0$, squashes the base manifold $\mb{C}P^2$. We consider a mass deformed M5 brane theory that we will call $(2,0)^*$ theory. Mass deformation breaks supersymmetry by half and the $(2,0)$ tensor multiplet splits into one $(1,0)$ tensor multiplet and one $(1,0)$ hypermultiplet with a mass parameter $m_H$. The R-symmetry is $SU(2)_R$ and the flavor symmetry is $SU(2)_F$. We have the following commuting charges: three Cartans $j_i$ of $SO(6)$, one Cartan $R_1+R_2$ of $SU(2)_R$, one Cartan $R_1-R_2$ of $SU(2)_F$. We can associate one chemical potential to each of these Cartan generators while preserving one supercharge $Q_{j_1 j_2 j_3}^{R_1 R_2} = Q_{---}^{--}$ with the $SO(6)$ Cartan charges $j_i=-1/2$, R charges $R_1=R_2=-1/2$ and scaling dimension $\Delta = 1/2$. The following operators commute with $Q_{---}^{--}$,
\bea
\O_1 &=& \Delta + \frac{1}{2}(R_1+R_2)\cr
\O_2 &=& R_1-R_2\cr
\O_3 &=& j_1-j_2\cr
\O_4 &=& j_2-j_3\cr
\O_5 &=& j_1+j_2+j_3 - 2 R_1 - R_2
\eea
In \cite{Bhattacharya:2008zy}, \cite{Kim:2012qf}, \cite{Lockhart:2012vp} the charge $\O_5$ that is associated with squashing of the Hopf fiber was not included. The existence of $\O_5$ was first noted in \cite{Kim:2012tr}, \cite{Kim:2013nva}, although there again only four independent charges were again included in the 6d superconformal index. But we can have five independent mutually commuting charges that commute with $Q_{---}^{--}$. In particular the generator $\O_5$ is crucial in this paper when we consider the theory on a lens space. We assign chemical potentials to each of these charges and define the superconformal index as
\bea
I(\beta,\omega,a,b,c,m_H) &=& \tr (-1)^F \omega^{\O_5} e^{-\beta (\O_1 + m_H \O_2 + a j_1 + b j_2 + c j_3)} 
\eea 
Generating functions associated to this index are \cite{Benvenuti:2016dcs}, \cite{Bak:2016vpi}
\ben
f_{tensor}(\beta,\omega,a,b,c) &=& \frac{e^{-3\beta}\omega^3 - e^{-2\beta} \omega^2 (e^{\beta a}+e^{\beta b}+e^{\beta c})}{(1-\omega e^{-\beta(1+a)})(1-\omega e^{-\beta(1+b)})(1-\omega e^{-\beta(1+c)})}\label{ten}\\
f_{hyper}(\beta,\omega,a,b,c,m_H) &=& \frac{e^{-\frac{3}{2}\beta} (\omega e^{\beta m_H}+\omega^2 e^{-\beta m_H})}{(1-\omega e^{-\beta(1+a)})(1-\omega e^{-\beta(1+b)})(1-\omega e^{-\beta(1+c)})}\label{hyp}
\een
for the tensor and hypermultiplets respectively. Here $m_H$ is a parameter that determines the hypermultiplet mass and we will put the radius of $S^5$ to be $r=1$. 

We will obtain the abelian indices on $\mb{R}\times (S^5/\mb{Z}_p)$ by Fourier transforming with respect to the chemical potential $\omega$. We do this first for the special case $a=b=c=0$ in section \ref{sectionunsquashed}, and later in full generality in section \ref{sectionsquashed} and obtain the results presented in eqs (\ref{generaltensor}) and (\ref{hyperconj}). We dualize the corresponding indices and obtain their high temperature expansions by using either one of three different dualization methods: zeta function regularization and the Abel-Plana formula method following \cite{Asorey:2012vp} for the unsquashed case $a=b=c=0$, and the plethystic dualization method following \cite{Kim:2012qf} for the squashed case with generic $a,b,c$. We then show that the unsquashed limit $a,b,c\rightarrow 0$ is both well-defined (independent of how we take the limit) and smooth. In the limit, we recover the previously obtained results for the unsquashed case with $a=b=c=0$. We summarize these three dualization methods in three appendices. Finally, in section \ref{sectS} we test an asymptotic S-duality conjecture \cite{Shaghoulian:2016gol} in the limit of large $p$.

\section{Supersymmetry enhancement}
We define the M5 brane generating function
\bea
f_{M5} &=& f_{tensor} + f_{hyper}
\eea
But for generic mass parameter $m_H$ this is a mass deformed version of the M5 brane. It has the same field content as the M5 brane, but not enough supersymmetry to make this a theory of a single $(2,0)$ tensor multiplet. We call this a $(2,0)^*$ theory. The preserved supercharges can be easily counted for each new chemical potential we insert into the index. Supercharges that preserve $\O_1$ have charges such that $R_1+R_2=-1$, hence they are $Q^{--}_{j_1 j_2 j_3}$. There are $8$ such supercharges. This is the amount of supersymmetry that we have in the $(2,0)^*$ theory for generic $m_H$ since these supercharges also commute with $\O_2$ whose chemical potential is $m_H$. But let us now consider the linear combination $\O_1 + m_H\O_2$ that appears in the index. The condition for this to commute with the supercharges, is that 
\bea
\frac{1}{2} + \(\frac{1}{2}+m_H\)R_1 + \(\frac{1}{2}-m_H\)R_2 &=& 0
\eea
For generic values of $m_H$ the only solution is $R_1 = R_2 = -1/2$. But for $m_H = 1/2$ we only need $R_1 = -1/2$ while $R_2$ can be either of $\pm 1/2$. This means we have enhancement of supersymmetry from $8$ to $16$ supercharges at $m_H = 1/2$. A similar enhancement of supersymmetry happens at $m_H = -1/2$ where instead $R_1$ can be either of $\pm 1/2$ and $R_2 = -1/2$. Let us now turn on $\O_5$. Then for $m_H = 1/2$ we find the four preserved supercharges, $Q^{--}_{---},Q^{-+}_{+--},Q^{-+}_{-+-},Q^{-+}_{--+}$ and for $m_H = -1/2$ we find the four preserved supercharges $Q^{--}_{---},Q^{+-}_{++-},Q^{+-}_{+-+},Q^{+-}_{-++}$. Had we instead defined $\O_5$ the symmetric way as
\bea
\O^{sym}_5 &=& j_1+j_2+j_3 - \frac{3}{2}(R_1 + R_2)
\eea
we would only get one preserved supercharge $Q^{--}_{---}$ when we turn on the chemical potential for $\O^{sym}_5$. We would like to preserve as much supersymmetry as possible for each new chemical potential that we turn on, and therefore we will not define $\O_5$ in the symmetric way. 

If we pick $m_H = 1/2$ and $a=b=c=0$, then we find a simplification also when we turn on the chemical potential $\omega$ (corresponding to $\O_5$). The generating function becomes 
\bea
f_{M5}(\beta,\omega) &=& \frac{\omega e^{-\beta}}{1-\omega e^{-\beta}}
\eea
This simplification would not occur had we instead used $\O_5^{sym}$, but also we do not get this simplification if we take $m_H = -1/2$ despite then again we have four supercharges. Instead, the simplification at $m_H = -1/2$ occurs if we replace $\O_5$ with $\O_5^{\vee} = j_1+j_2+j_3-R_1-2R_2$. We normalized $\O_5$ such that $e^{2\pi i \O_5}=1$ on bosonic states. This gives the quantization condition $\O_5 \in \mb{Z}$. By lensing, we will identify the rotation by the angle $2\pi/p$ with the identity operator, and this gives the quantization condition $\O_5 \in p \mb{Z}$. To project out all the integer modes except multiples of $p$, we put $\eta = e^{2\pi i/p}$ and introduce the projection operator
\bea
\frac{1}{p} \sum_{\l=0}^{p-1} \eta^{\l} &=& \sum_{\l\in \mb{Z}} \delta_{n,\l p}
\eea
Thus we get the generating function on $\mb{R}\times (S^5/{\mb{Z}_p})$ as
\bea
f_{M5}(\beta,p) &=& \frac{1}{p} \sum_{\l=0}^{p-1} f_{M5}(\beta,\eta^{\l})
\eea
Lensing by using the generator $\O_5$ leads to the lens space $L(p;1,1)$. For the lens space $S^5/{\mb{Z}_p} = L(p;q_1,q_2)$ we shall replace $\O_5$ with the operator
\bea
\O_{q_1,q_2} &=& q_1 j_1 + q_2 j_2 + j_3 - (q_1+q_2)R_1 - R_2
\eea
Then for $m_H = 1/2$ we have generically only two conserved supercharges $Q^{--}_{---},Q^{--}_{--+}$, and for $m_H = -1/2$ we preserve $Q^{--}_{---},Q^{+-}_{++-}$. It is easy to compute the generating function on $\mb{R}\times L(p;1,1)$ for $m_H = 1/2$. The result is
\bea
f_{M5}(\beta,p) &=& \frac{e^{-p\beta}}{1-e^{-p\beta}}
\eea
Lensing for the $(1,0)$ tensor multiplet does not depend on the choice of $m_H$ nor on the detailed definition of $\O_5$, whether we use $\O_5,\O_5^{\vee},\O_5^{sym}$ or some other combination. This is simply because all fields in the $(1,0)$ tensor multiplet are neutral with respect to the R-charges $R_1,R_2$. But for the hypermultiplet, lensing will depend on the choice of $\O_5$. Does this mean that lensing leads to an ambigous result for the hypermultiplet? We will now explain that the answer is either yes or no depending on the amount of supersymmetry that we like to preserve. When we put a theory on a curved manifold, we always need to specify the amount of supersymmetry that we like to preserve. At the point where $a=b=c=0$ and $m_H=1/2$, the preserved supercharges may tell us how we shall choose $\O_5$ so that there is no ambiguity in the lensing procedure. In this paper we will consider the $(2,0)^*$ theory that has four preserved supercharges $Q^{--}_{---},Q^{-+}_{+--},Q^{-+}_{-+-},Q^{-+}_{--+}$ at the point specified by $a=b=c=0$ and $m_H=1/2$. This specification uniquely tells us that we shall use the operator $\O_5$ to define the superconformal index to squash the Hopf fiber, rather than $\O_5^{\vee}$ or some other combination. If we would like to preserve less than four supercharges at this particular point, then we will find a certain ambiguity in how to pick $\O_5$. But this is a common situation. When we lower the amount of supersymmetry we always find more freedom in how to set up a corresponding theory that preserves that amount of supersymmetry.

\section{From generating function to superconformal index}
If we have a spectrum of discrete energy levels, then this may be encoded in a generating function
\bea
f(\beta) &=& \sum_n d_n e^{-\beta E_n}
\eea
where $E_n$ are the energy levels with degeneracies $d_n$ and $\beta$ is the inverse temperature. (To get the generating function of an index, we count degeneracies of fermionic fields with a minus sign.) The index is an infinite product 
\bea
I(\beta) &=& \prod_{\l\in\mb{Z}} \prod_n (a_{\l}^2 + E_n^2)^{-d_n/2}
\eea
where 
\bea
a_{\l} &=& \frac{2\pi \l}{\beta}
\eea
The product is divergent and needs to be regularized. We may use zeta function regularization and define \cite{Asorey:2012vp}
\ben
\zeta(s) &=& \mu^{2s} \sum_n \sum_{\l\in\mb{Z}} d_n (a_{\l}^2+E_n^2)^{-s}\label{zeta0}
\een
where $\mu$ is an energy scale that we need to insert in order for $\zeta(s)$ to be dimensionless. Then the index will be given by 
\ben
I(\beta) = e^{-\beta F(\beta)} = e^{\frac{1}{2}\zeta'(0)}\label{defzeta}
\een
In \cite{Asorey:2012vp} it was shown by a direct computation that (\ref{defzeta}) can be rewritten as a low temperature expansion 
\ben
I(\beta) &=& e^{-\beta E} \prod_n (1-e^{-\beta E_n})^{-d_n}\label{lowT}
\een
where $E$ is the Casimir energy. The nice feature with the derivation in \cite{Asorey:2012vp}, is that the Casimir energy factor appears automatically and does not need to be multiplied by hand. If we take the logarithm, we get
\bea
\log I(\beta) &=& -\beta E - \sum_n d_n \log(1-e^{-\beta E_n})\cr
&=& -\beta E + \sum_n \sum_{k=1}^{\infty} \frac{1}{k} d_n e^{-k\beta E_n}
\eea
If we exchange the sums, the result can be expressed as the plethystic sum of the generating function
\bea
\log I(\beta) &=& -\beta E + \sum_{k=1}^{\infty} \frac{1}{k} f(k\beta)
\eea
The plethystic sum may involve a divergent sum in certain applications. Then, even if this divergent sum is canceled out in the dualization process that takes us from the low temperature expansion (\ref{lowT}) to a high temperature expansion, we may nevertheless miss out some logarithmic terms (or prefactors of the superconformal index). Another drawback is that the plethystic method can only be applied to generating functions $f(\beta)$ that are antisymmetric, $f(\beta) = - f(-\beta)$. The advantage with the plethystic method is that it leads to simpler computations and in particular it is easy to extract the Stefan-Boltzmann terms from the generating function, which is much more involved to do by using the other methods.

\section{Turning off squashing parameters}\label{sectionunsquashed}
In this section we put the squashing parameters to zero, $a=b=c=0$, and obtain the generating functions on $S^1 \times (S^5/{\mb{Z}_p})$ by Fourier transforming with respect to the chemical potential $\omega$. We also obtain the high temperature expansions of the corresponding indices by using zeta function dualization method, the Abel-Plana dualization method, and the plethystic dualization method. We outline each of these dualization methods in three appendices. 

\subsection{The $(1,0)$ tensor multiplet index}
The generating function for the $(1,0)$ tensor multiplet on $\mb{R}\times S^5$ is given by
\bea
f_{tensor}(\beta,\omega) &=& \frac{\omega^3 e^{-3\beta}-3 \omega^2 e^{-2\beta}}{(1-\omega e^{-\beta})^3}
\eea
where we just turn on one chemical potential $\omega = e^{-\beta h}$ that corresponds to a rotation along the Hopf fiber (and $h$ was introduced in the Introduction). By noting that 
\bea
\frac{1}{(1-e^{-\beta})^3} &=& \sum_{n=0}^{\infty} \frac{(n+1)(n+2)}{2} e^{-\beta n}
\eea
we can write this as
\bea
f_{tensor}(\beta,\omega) &=& \sum_{n=1}^{\infty} (1-n^2) \omega^n e^{-\beta n}
\eea
To obtain the generating function on the lens space $S^5/\mb{Z}_p = L(p;1,1)$, we sum over all $\omega$ that are taken as the $p$ different $p$-th roots of unity. This sum amounts to projecting onto the lens space where we are identifying
\bea
\psi &\sim & \psi + \frac{2\pi}{p}
\eea
where $\psi$ is the $2\pi$ ranged Hopf fiber coordinate on $S^5$. We then use the identity
\bea
\frac{1}{p}\sum_{\l=0}^{p-1} e^{2\pi i \l n/p} &=& \sum_{q\in\mb{Z}} \delta_{n,pq}
\eea
and get
\bea
f_{tensor}(\beta,p) &=& \sum_{n=1}^{\infty} (1-n^2 p^2) e^{-\beta p n}
\eea
which we evaluate to 
\ben
f_{tensor}(\beta,p) &=& \frac{e^{-p\beta}}{1-e^{-p\beta}} - \frac{p^2}{4} \frac{\cosh\frac{p\beta}{2}}{\(\sinh\frac{p\beta}{2}\)^3}\label{gentensor0}
\een
The small-$\beta'$ expansion is
\ben
f_{tensor}(\beta',p) &=& -\frac{2}{p\beta'^3} + \frac{1}{p \beta'} - \frac{1}{2} + \frac{p^3+10p}{120}\beta' + \O(\beta'^2)\label{tensorsmall}
\een
The low temperature (or large $\beta$) behavior of the full index is governed by the Casimir energy $E$, 
\bea
I_{tensor}(\beta,p) &\sim& e^{-\beta E}
\eea
The Casimir energy can be read off from the small-$\beta'$ expansion of the generating function, by interpreting $\beta'$ as a small regulator, and after subtracting the divergent terms proportional to $\beta'^{-3}$ and $\beta'^{-1}$, thus defining a renormalized $f_{ren}$. The Casimir energy is then computed as 
\bea
E &=& -\frac{1}{2} \frac{d}{d\beta'} f_{tensor,ren}(\beta',p)\Bigg|_{\beta'=0}
\eea
We get
\bea
E_{tensor} &=& -\frac{p^3+10p}{240}
\eea
If we use zeta function regularization, we will compute the Casimir energy as follows. We begin by defining
\bea
E(s) &=& \frac{1}{2} \sum_{n=1}^{\infty} (1-p^2n^2) (pn)^{1-s}\cr
&=& \frac{1}{2} p^{1-2s} \(\zeta(s-1) - p^2 \zeta(s-3)\)
\eea
and then 
\bea
E(0) &=& \frac{1}{2} p \(\zeta(-1) - p^2 \zeta(-3)\)\cr
&=& -\frac{p^3 + 10 p}{240}
\eea
is the Casimir energy. The advantage with this latter method is that we do not need to worry about how to remove the singular terms proportional to $\beta'^{-3}$ and $\beta'^{-1}$.

The zeta function that is associated to the generating function $f_{tensor}(\beta,p)$ is
\ben
\zeta(s) &=& \mu^{2s} \sum_{\l\in \mb{Z}} \sum_{n=1}^{\infty} (1-n^2p^2)(n^2p^2+a_{\l}^2)^{-s}\label{tzeta}
\een
We decompose $\zeta(s) = \zeta_{\l=0}(s) + \sum_{\l\neq 0} \zeta_{\l\neq 0}(s)$ and rewrite
\bea
\zeta_{\l=0}(s) &=& \mu^{2s} \(p^{-2s}\zeta(2s)-p^{2-2s} \zeta(2s-2)\)\cr
\zeta_{\l\neq 0}(s) &=& \mu^{2s} \sum_{n=1}^{\infty} \((1+a_{\l}^2)(n^2p^2+a_{\l}^2)^{-s} - (n^2p^2+a_{\l}^2)^{1-s}\)
\eea
Then by using $\zeta(0) = -\frac{1}{2}$, $\zeta(-2) = 0$ and $\zeta'(0) = -\frac{1}{2}\log(2\pi)$, $\zeta'(-2) = -\frac{\zeta(3)}{4\pi^2}$ we get
\bea
\zeta'_{\l=0}(0) &=& - \log\frac{\mu}{p} - \log(2\pi) + p^2 \frac{\zeta(3)}{2\pi^2}
\eea
For $\l\neq 0$ we get\footnote{We have a sum over $n=1,2,...$ that we wish to dualize. To this end, we add the term that we would get by taking $n=0$. We also add the terms with $n$ replaced by $-n$, which is possible since only $n^2$ appears in the summand. The details are in appendix $A$.}
\bea
\zeta'_{\l\neq 0,n=0}(0) &=& -\frac{2\pi^4}{45p\beta^3}+\frac{\pi^2}{3p\beta}+\log(\mu \beta)
\eea
By also including the sectors with $n\neq 0$, we get the result
\ben
-\beta F_{tensor} &=& -\frac{\pi^4}{45p\beta^3} + \frac{\pi^2}{6p\beta} + \frac{1}{2}\log \frac{p\beta}{2\pi} + p^2\frac{\zeta(3)}{4\pi^2}\cr
&& + \sum_{\l=1}^{\infty} \sum_{n=1}^{\infty} \frac{p^2}{n^3} \(\frac{1}{2\pi^2}+ \frac{2\l n}{p\beta} + \frac{n^2}{p^2} + \(\frac{2\pi \l n}{p\beta}\)^2\) e^{-\frac{4\pi^2 \l n}{p\beta}}\label{unsquashedtensor}
\een
In particular, we see that the dependence on the scale $\mu$ cancels out. We computed the contribution from terms with $n\neq 0$ by using the zeta function dualization method that we outline in appendix $A$, following \cite{Asorey:2012vp}.

\subsection{The $(1,0)$ hypermultiplet index}
If we pick the hypermultiplet mass parameter $m_H = 1/2$, then we have the following generating function 
\bea
f_{hyper}(\beta) &=& \frac{e^{-2\beta}+e^{-\beta}}{(1-e^{-\beta})^3}
\eea
We may use 
\bea
\frac{1}{(1-e^{-\beta})^3} &=& \sum_{n=0}^{\infty} \frac{1}{2}(n+1)(n+2) e^{-\beta n}
\eea
to bring this into the form
\bea
f_{hyper}(\beta) &=& \sum_{n=1}^{\infty} n^2 e^{-\beta n}
\eea
Now let us take $m_H = 1/2 + \epsilon$. Then the generating function is
\bea
f_{hyper}(\beta) &=& \frac{1}{2}\sum_{n=0}^{\infty} (n-1)n e^{-\beta(n+\epsilon)}+\frac{1}{2}\sum_{n=0}^{\infty} (n+1)n e^{-\beta(n-\epsilon)}
\eea
We will now use the Abel-Plana method to obtain the high temperature expansion. We outline this method in the appendix. Associated to this is the function
\bea
f(z) &=& \frac{1}{2}(z-1)z \log(1-e^{-\beta(z+\epsilon)}) + \frac{1}{2}(z+1)z \log(1-e^{-\beta(z-\epsilon)})
\eea
Now we Taylor expand this function in $\epsilon$
\bea
f(z) &=& z^2 \log(1-e^{-\beta z}) - \beta \epsilon \frac{z e^{-\beta z}}{1-e^{-\beta z}} - \frac{1}{2}\beta^2 \epsilon^2 \frac{z^2 e^{-\beta z}}{(1-e^{-\beta})^2} + \O(\epsilon^3)
\eea
We have 
\bea
\int_0^{\infty} dx f(x) &=& -\frac{\pi^4}{45 \beta^3} - \epsilon \frac{\pi^2}{6\beta} - \epsilon^2 \frac{\pi^2}{6\beta} + \O(\epsilon^3)
\eea
which is consistent with 
\bea
\frac{\pi^4}{45\beta^3} + \frac{12m_H^2-3}{12} \frac{\pi^2}{6\beta}
\eea
by taking $m_H = 1/2+\epsilon$ and expanding in $\epsilon$. A rather curious fact is that the series terminates at order $\epsilon^2$. But the integrals at order $\epsilon^n$ for $n>2$ all diverges, although by summing them all up, we shall of course expect a finite result, and actually the contribution from all those divergent integrals should sum up to zero. 

It is easy to get the Casimir energy from the generating function. We get
\bea
f(\beta',m_H) &=& \frac{2}{\beta'^3} + \frac{4m_H^2-1}{4\beta'} + \frac{1}{960}(17-120m_H^2+80m_H^4) \beta' + \O(\beta'^2)
\eea
Then we see that the Casimir energy is
\bea
E &=& \frac{1}{240} + \epsilon\frac{1}{24} - \epsilon^3 \frac{1}{12} - \epsilon^4 \frac{1}{24}
\eea
as an exact expression in $\epsilon$.

\subsubsection{Refinement}
If we take $m_H = 1/2 + \epsilon$, then we have
\bea
f_{hyper} &=& \frac{1}{2}\sum_{n=0}^{\infty} (n+1)(n+2) \(e^{-\beta (n+2+\epsilon)} \omega^{n+2} + e^{-\beta(n+1-\epsilon)} \omega^{n+1}\)
\eea
Now we can lens this expression, and get
\ben
f_{hyper} &=& \frac{1}{2}\sum_{q=0}^{\infty} (pq-1)pq e^{-\beta(pq+\epsilon)} + \frac{1}{2} \sum_{q=0}^{\infty} (pq+1)pq e^{-\beta(pq-\epsilon)}\cr
&=& \frac{p^2 \cosh(\beta \epsilon)\cosh\frac{p\beta}{2} + p\sinh(\beta\epsilon) \sinh\frac{p\beta}{2}}{4 \(\sinh\frac{p\beta}{2}\)^3}\label{hypprevious}
\een
From this we can extract the Casimir energy
\bea
E_{hyper} &=& \frac{p^3}{240} + \epsilon \frac{p}{24} - \epsilon^3 \frac{1}{12 p} -\epsilon^4 \frac{1}{24p}
\eea
For the high temperature expansion, we define
\bea
f(z,\epsilon) &=& \frac{1}{2}\Big((p z - 1) p z \log\(1 - e^{-\beta (p z + \epsilon)}\) + (p z + 1) p z \log\(1 - e^{-\beta (p z - \epsilon)}\)\Big)
\eea
that we Taylor expand in $\epsilon$ and then we are able to compute its integral to the first few orders, with the result 
\bea
\int_0^{\infty} dx f(x,\epsilon) &=& -\frac{\pi^4}{45p\beta^3}-\epsilon\frac{\pi^2}{6p\beta} - \epsilon^2 \frac{\pi^2}{6p\beta} + \O(\epsilon^3)
\eea
Let us move on to the other term in the Abel-Plana formula
\bea
i \int_0^{\infty} dx \frac{f(ix)-f(-ix)}{e^{2\pi x}-1}
\eea
We will content ourselves to compute up to linear order in $\epsilon$, and thus we only need to work with $f$ expanded to first order
\bea
f(z,\epsilon) &=& p^2 z^2 \log(1-e^{-p\beta z}) + \epsilon p\beta \frac{z e^{-p\beta z}}{1-e^{-p\beta z}}
\eea
To this order, we get
\bea
f(ix,\epsilon) - f(-ix,\epsilon) &=& p^2 x^2\log\(-e^{i p \beta x}\) + i p x \beta \epsilon\cr
&=& p^2 x^2 i \(p \beta x - 2\pi n - \pi\)  + i p x \beta \epsilon
\eea
where 
\bea
\int_0^{\infty} dx &=& \sum_{n=0}^{\infty} \int_{2\pi n}^{2\pi(n+1)} dx
\eea
Thus we like to compute (and here we changed the sign of $f(z)$)
\bea
\sum_{n=0}^{\infty} \int_{2\pi n/(p\beta)}^{2\pi(n+1)/(p\beta)} dx \frac{p^2 x^2 \(p \beta x - 2\pi n - \pi\)  + p x \beta \epsilon}{e^{2\pi x}-1}
\eea
We report the following partial results,
\bea
\int_0^{\infty} dx \frac{x^3}{e^{2\pi x}-1} &=& \frac{1}{240}\cr
\int_0^{\infty} dx \frac{x^2}{e^{2\pi x}-1} &=& \frac{\zeta(3)}{4\pi^3}\cr
\int_0^{\infty} dx \frac{x}{e^{2\pi x}-1} &=& \frac{1}{24}
\eea
We may expand the denominator as
\bea
\frac{e^{-2\pi x}}{1-e^{-2\pi x}} &=& \sum_{k=1}^{\infty} e^{-2\pi k x}
\eea
and consider the integrals
\bea
\int dx x^2 e^{-2\pi k x} &=& -\frac{1}{4\pi^3 k^3} (1+2\pi k x + 2\pi^2 k^2 x^2) e^{-2\pi k x}
\eea
We get
\bea
\frac{p^3 + 10 \epsilon p}{240}\beta - p^2\frac{\zeta(3)}{4\pi^2} 
\eea
plus the double sum
\bea
&&\frac{p}{2\pi^2} \sum_{k,n=1}^{\infty} \frac{n}{k^3} \[\(1+2\pi k x + 2\pi^2 k^2 x^2\) e^{-2\pi k x}\]_{2\pi n/(p\beta)}^{2\pi(n+1)/(p\beta)}\cr
&=& - \sum_{k,n=1}^{\infty} \frac{p^2}{k^3} \(\frac{1}{2\pi^2}+\frac{2kn}{p\beta}+\(\frac{2\pi k n}{p\beta}\)^2\) e^{-\frac{4\pi^2 k n}{p\beta}}
\eea
Upon adding the Casimir energy term $-\beta E_{hyper}$ we find a cancelation.
Summarizing, we have got
\bea
-\beta F_{hyper} &=& \frac{\pi^4}{45p\beta^3}+\epsilon\frac{\pi^2}{6p\beta}+\epsilon^2 \frac{\pi^2}{6p\beta}-p^2\frac{\zeta(3)}{4\pi^2}\cr
&&- \sum_{k,n=1}^{\infty} \frac{p^2}{k^3} \(\frac{1}{2\pi^2}+\frac{2kn}{p\beta}+\(\frac{2\pi k n}{p\beta}\)^2\) e^{-\frac{4\pi^2 k n}{p\beta}}
\eea

If we add the contributions of the tensor and the hypermultiplets, we get
\bea
-\beta \(F_{tensor} + F_{hyper}\) &=& \frac{\pi^2(1+\epsilon+\epsilon^2)}{6p\beta} + \frac{1}{2} \log \frac{p\beta}{2\pi} + \sum_{n,k=1}^{\infty} \frac{1}{n} e^{-\frac{4\pi^2 k n}{p\beta}}
\eea
If we put $\epsilon = 0$, we may express this result as 
\bea
I(\beta) &=& \sqrt{\frac{p\beta}{2\pi}} e^{\frac{\pi^2}{6p\beta}} \exp\(\sum_{n=1}^{\infty}\frac{1}{n}\frac{e^{-\frac{4\pi^2 n}{p\beta}}}{1-e^{-\frac{4\pi^2 n}{p\beta}}}\)
\eea
This result was also obtained in \cite{Kim:2012qf} by using the known modular property of the Dedekind eta function. 

We can also easily perform the sum over $n$. For the hypermultiplet and if we put $\epsilon = 0$, we then get
\bea
-\beta F_{hyper} &=& \frac{\pi^4}{45p\beta^3} - \sum_{k=1}^{\infty} \[\frac{p^2}{4\pi^2 k^3} \frac{\cosh \frac{2\pi^2 k}{p\beta}}{\sinh \frac{2\pi^2 k}{p\beta}} + \frac{p}{2 k^2\beta} \frac{1}{\(\sinh \frac{2\pi^2 k}{p\beta}\)^2} + \frac{\pi^2}{k\beta^2} \frac{\cosh \frac{2\pi^2 k}{p\beta}}{\(\sinh \frac{2\pi^2 k}{p\beta}\)^3}\]
\eea

\subsubsection{The same result from the plethystic exponent}
Let us return to the generating function (\ref{hypprevious}) and take $\epsilon = 0$ for simplicity,
\ben
f_{hyper}(\beta,p) &=& \frac{p^2}{4} \frac{\cosh\frac{p\beta}{2}}{\(\sinh \frac{p\beta}{2}\)^3}\cr
&=& \frac{2}{p\beta^3} - \frac{p^3}{120}\beta + \O(\beta^2)\label{hypersmall}
\een
We can use this to compute the index by taking the plethystic exponent,
\bea
\log \t I_{hyper}(\beta,p) &=& \sum_{n=1}^{\infty} \frac{1}{n} f_{hyper}(n\beta,p)\cr
&=& \sum_{n=-\infty}^{\infty} \int_{\epsilon}^{\infty} \frac{ds}{s} f_{hyper}(s\beta,p) e^{2\pi i n s}
\eea
The correct index is given by $I_{hyper} = e^{-\beta E} \t I_{hyper}$. For the second identity to hold, we need to pick up the points $s=n$ for $n=1,2,...$ from
\bea
\sum_{n\in \mb{Z}} e^{2\pi i n s} = \sum_{n\in \mb{Z}} \delta(s-n)
\eea
To this end, we shall take lower integration bound such that $0<\epsilon<\beta$. 

Dualization is now performed as follows. First we regularize as follows
\bea
f(s) &=& f_{sing}(s) + f_{reg}(s)
\eea
where $f_{reg}(s) = f(s) - f_{sing}(s)$ and $f_{sing}(s)$ involves terms of the form $s^{-n}$ for $n>0$ such that $f_{reg}(0)$ is finite. Next we can compute the contribution from $f_{sing}(s)$. We have the small-$s$ expansion
\bea
f_{hyper}(s\beta,p) &=& \frac{2}{p \beta^3 s^3} - \frac{p^3 \beta s}{120} + \O(s^3)
\eea
The first term in this expansion is the singular term, which gives rise to the Stefan-Boltzmann term
\bea
\sum_{n=-\infty}^{\infty} \int_{\epsilon}^{\infty} ds \frac{2}{p \beta^3 s^4} e^{2\pi i n s} = \frac{2}{p\beta^3}\zeta(4) = \frac{\pi^4}{45 p \beta^3}
\eea
The remaining piece is 
\bea
\sum_{n=1}^{\infty} \int_{-\infty}^{\infty} \frac{ds}{s} f_{reg}(\beta s) e^{2\pi i n s} + \frac{1}{2} \int_{-\infty}^{\infty} \frac{ds}{s} f_{reg}(\beta s)  - \frac{1}{2} \frac{f_{reg}(s)}{s}|_{s=0}
\eea
There are triple poles at 
\bea
s &=& \frac{2\pi i k}{p\beta}, \qquad k\in{\mb{Z}}
\eea
and by encircling those triple poles that lie in the upper half plane and picking up the residues, we get  
\ben
\sum_{n=1}^{\infty} \int_{-\infty}^{\infty} \frac{ds}{s} f_{reg}(\beta s) e^{2\pi i n s}
&=& -\sum_{n=1}^{\infty} \sum_{k=1}^{\infty} \frac{p^2}{k^3} \(\frac{1}{2\pi^2}+ \frac{2k n}{p\beta} + \(\frac{2\pi k n}{p\beta}\)^2\) e^{-\frac{4\pi^2 k n}{p\beta}}\label{hyperinst}\\
\frac{1}{2}\int_{-\infty}^{\infty} \frac{ds}{s} f_{reg}(\beta s) &=& -\frac{p^2}{4\pi^2} \zeta(3)\label{hyperpert}
\een
By a general argument that we prove in the appendix, we have that
\bea
- \frac{1}{2} \frac{f_{reg}(s)}{s}|_{s=0}
\eea
is canceled by the Casimir energy.

By comparing the two computations, and by looking at the term in (\ref{hyperpert}), we see that what was refered to as the topological subleading term in \cite{Asorey:2012vp} corresponds to what was refered to as the perturbative contribution in \cite{Kim:2012qf}.

\section{Turning on squashing parameters}\label{sectionsquashed}
We will now consider a more complicated situation with generic squashing parameters turned on. For this case, the plethystic method is easy to use, whereas the other two methods, the generalized zeta function and the Abel-Plana formula, become difficult to use. The generating functions for general squashing parameters $a,b,c$ and generic lensing parameter $p$, are quite complicated. However, already from the small $\beta$ expansion of these generating functions, we can extract the high temperature and low temperature asymptotic behavior of the corresponding indices, so that is where we will start.

We begin by extracting the Stefan-Boltzmann terms with squashing. Given a generating function $f(\beta)$, the Stefan-Boltzmann terms are obtained by computing the following quantity,
\bea
\sum_{n\in\mb{Z}} \int_{\epsilon}^{\infty} \frac{ds}{s} e^{\frac{2\pi i n s}{\beta}} f_{sing}(s)
\eea
For the explicit computations, we will only need the following results,
\bea
\sum_{n\in\mb{Z}} \int_{\epsilon}^{\infty} \frac{ds}{s} e^{\frac{2\pi i n s}{\beta}} \frac{1}{s^n} &=& \frac{\zeta(n+1)}{\beta^n}
\eea
and in this paper we will encounter
\bea
\zeta(4) &=& \frac{\pi^4}{90}\cr
\zeta(2) &=& \frac{\pi^2}{6}
\eea

\subsection{The tensor multiplet}
The refined generating function for the tensor multiplet is given by (\ref{ten}). We begin by writing a series expansion for
\ben
\frac{1}{\(1-\omega e^{-\beta(1+a)}\)\(1-\omega e^{-\beta(1+b)}\)\(1-\omega e^{-\beta(1+c)}\)} &=& -\sum_{n=0}^{\infty} \omega^n e^{-\beta n} f_n(a,b,c)\label{series}
\een
where
\bea
f_n(a,b,c) &=& \frac{e^{-\beta a (n+1)}(e^{\beta b}-e^{\beta c}) + e^{-\beta b (n+1)}(e^{\beta c}-e^{\beta a}) + e^{-\beta c (n+1)}(e^{\beta a}-e^{\beta b})}  
{(1-e^{-\beta(a-b)})(1-e^{-\beta(b-c)})(1-e^{-\beta(c-a)})}
\eea
We now need to multiply this by the numerator
\bea
\omega^3 e^{-3\beta} - \omega^2 e^{-2\beta} (e^{\beta a} + e^{\beta b} + e^{\beta c})
\eea
We then get two terms, $\omega^{n+3}(...) + \omega^{n+2}(...)$. The trick is to shift the sum for the first term to bring both terms into the same form $\omega^{n+2}(...+...)$, and then replace $n+2$ by $mp$ where $m=1,2,...$ for $p>1$. (If $p=1$, then we shall take $m=2,3,...$.) For $p>1$ we get
\bea
f_{tensor}(p,\beta,a,b,c) &=& -\sum_{m=1}^{\infty} e^{-\beta p m} \(f_{p m - 3}(\beta,a,b,c) - (e^{\beta a}+e^{\beta b} + e^{\beta c}) f_{p m - 2}(\beta,a,b,c)\)
\eea
This is a geometric sum, which we evaluate to
\ben
f_{tensor}(p,\beta,a,b,c) &=& -\frac{1+e^{\beta (b-c)}}{(1-e^{-\beta(a-b)})(1-e^{-\beta(c-a)})} \frac{1}{1-e^{p\beta(1+a)}}+\cycl\label{gentensor}
\een
We note that 
\bea
f_{tensor}(p,\beta,a,b,c) + f_{tensor}(p,-\beta,a,b,c) &=& -1
\eea
This enables us to write this in the manifestly antisymmetric form by adding $-1/2$,
\bea
f_{tensor}(p,\beta,a,b,c) &=& \(\frac{\cosh\frac{\beta}{2}(b-c)}{4\sinh\frac{\beta}{2}(a-b)\sinh\frac{\beta}{2}(a-c)} \frac{\cosh\frac{p\beta}{2}(1+a)}{\sinh\frac{p\beta}{2} (1+a)}+\cycl\)-\frac{1}{2}
\eea
The small $\beta$ expansion reads
\bea
f_{tensor}(p,\beta,a,b,c) &=& -\frac{2}{Np \beta^3}+\frac{6+5(ab+bc+ca)}{6Np\beta}-\frac{1}{2}\cr
&+& \frac{p^3 + 10p}{120}\beta - \frac{a^4+b^4+c^4-36abc}{240Np}\beta + \O(\beta^2)
\eea
where $N := (1+a)(1+b)(1+c)$ and is a generalization of (\ref{tensorsmall}) to the squashed case. From the singular terms we extract the Stefan-Boltzmann terms
\bea
\beta F_{tensor} &=& \frac{\pi^4}{45 Np \beta^3}-\frac{(6+5(ab+bc+ca))\pi^2}{36Np\beta}
\eea
and from the linear term we extract the Casimir energy
\bea
E_{tensor} &=& -\frac{p^3 + 10p}{240} + \frac{a^4+b^4+c^4-36abc}{480 N p}
\eea

\subsection{The hypermultiplet}
We pick $m_H = 1/2+\epsilon$ and put $t=e^{\beta \epsilon}$ where we have the following refined generating function,
\bea
f_{hyper}(\beta,\omega,t) &=& \frac{e^{-2\beta}\omega^2 t^{-1} +e^{-\beta}\omega t}{(1-\omega e^{-\beta(1+a)})(1-\omega e^{-\beta(1+b)})(1-\omega e^{-\beta(1+c)})}
\eea
We again use the series expansion (\ref{series}) for the denominator and write
\bea
f_{hyper}(\beta,\omega) &=& - \sum_{n=0}^{\infty} \omega^n e^{-\beta n} (e^{-2\beta}\omega^2 t^{-1}+e^{-\beta}\omega t) f_n(a,b,c)\cr
&=& -\sum_{n=-1}^{\infty} \omega^{n+2} e^{-\beta(n+2)}(f_n t+f_{n+1}t^{-1})
\eea
Summing over $\omega$ picks out $n+2=mp$ for $m=1,2,...$. We get
\bea
f_{hyper}(\beta,p,a,b,c,t) &=& \sum_{m=1}^{\infty} e^{-\beta m p} \(t f_{m p -1}+t^{-1}f_{m p-2}\)
\eea
The sum can be evaluated with the result
\ben
f_{hyper}(\beta,p,a,b,c,t) &=& \frac{t e^{-\beta c}+t^{-1} e^{\beta b}}{(1-e^{-\beta(a-b)})(1-e^{-\beta(c-a)})}\frac{1}{1-e^{p\beta(1+a)}} + \cycl\label{genhyper1}
\een
By noting that 
\bea
f_{hyper}(p,\beta,a,b,c,t) + f_{hyper}(p,-\beta,a,b,c,1/t) &=& 0
\eea
we can write this as
\bea
f_{hyper}(p,\beta,a,b,c,t) &=& \frac{\cosh\frac{\beta}{2}(b+c+2\epsilon)}{4\sinh\frac{\beta}{2}(a-b)\sinh\frac{\beta}{2}(a-c)}\frac{\cosh\frac{p\beta}{2}(1+a)}{\sinh\frac{p\beta}{2}(1+a)} + \cycl\label{genhyper2}
\eea
The small $\beta$ expansion reads
\ben
f_{hyper} &=& \frac{2}{Np\beta^3} + \frac{ab+bc+ca+6(\epsilon+\epsilon^2)}{6Np\beta} + \frac{M\beta}{120Np} + \O(\epsilon^2,\beta^2)\label{hypexpand}
\een
where
\bea
M &=& \frac{1}{2}\(a^4 + b^4 + c^4\) + 2 abc -N p^4\cr
&& + \(20 (ab+bc+ca) + 10abc - 10N p^2\)\epsilon\cr
&& +10 (ab+bc+ca) \epsilon^2 + 20 \epsilon^3 + 10 \epsilon^4
\eea
which is a generalization of (\ref{hypersmall}) to the squashed case. From the singular terms we extract the Stefan-Boltzmann terms
\bea
-\beta F_{hyper} &=& \frac{\pi^4}{45 Np\beta^3} + \frac{(ab+bc+ca+6(\epsilon+\epsilon^2))\pi^2}{36 N p\beta}
\eea
and from the linear term we extract the Casimir energy
\bea
E_{hyper} &=& \frac{p^3}{240} + \frac{p\epsilon}{24} - \frac{\epsilon^3}{12 N p} - \frac{\epsilon^4}{24 N p}\cr
&& - \frac{a^4+b^4+c^4+4abc}{480 N p} \cr
&& - \frac{2(ab+bc+ca)+abc}{24Np}\epsilon\cr
&& -\frac{ab+bc+ca}{24Np}\epsilon^2
\eea

\subsection{Summing the contributions}
If we sum the contributions from tensor and hypermultiplets,
\bea
f_{M5}(\beta,p) = f_{tensor}(\beta,p)+f_{hyper}(\beta,p)
\eea
we get the result
\ben
f_{M5} &=& \frac{\sinh \frac{\beta(\epsilon-c)}{2}\sinh \frac{\beta(\epsilon-b)}{2}}{\sinh \frac{\beta(a-b)}{2} \sinh \frac{\beta(c-a)}{2}} \frac{1}{1-e^{p\beta (1+a)}} + \cycl\label{Kimyeong}
\een
This result agrees precisely with eq (4.5) in \cite{Kim:2013nva}. At $\epsilon = 0$, the small $\beta$ expansion reads
\bea
f_{M5} &=& \frac{1+a b+a c+b c}{N p \beta} - \frac{1}{2} + \frac{1}{12}p\beta + \frac{abc}{6 Np} \beta + \O(\beta^2)
\eea
From the divergent term we obtain the Stefan-Boltzmann term
\bea
\log I_{SB} &=& \frac{1+ab+bc+ca}{N p} \frac{\pi^2}{6\beta}
\eea
and from the linear term we obtain the Casimir energy,
\bea
E_{M5} &=& -\frac{p}{24} - \frac{abc}{12 N p}
\eea

\subsection{More on the exact results}
The exact expressions for generating functions that we have obtained in (\ref{gentensor}) and (\ref{genhyper1}) have not yet been written in the fully reduced form, by which we mean the following. If we write the sum of the three cyclic permutations on a common denominator (for $i=\{tensor,hyper\}$ respectively)
\bea
f_i &=& \frac{P_i}{\(1-e^{\beta(b-a)}\)\(1-e^{\beta(c-b)}\)\(1-e^{\beta(a-c)}\)\(1-e^{p\beta(1+a)}\)\(1-e^{p\beta(1+b)}\)\(1-e^{p\beta(1+c)}\)}
\eea
then this can be always further reduced to the form
\bea
f_i &=& \frac{P_{i,reduced}}{\(1-e^{p\beta(1+a)}\)\(1-e^{p\beta(1+b)}\)\(1-e^{p\beta(1+c)}\)}
\eea
by which we mean that the poles associated to the vanishing of $1-e^{\beta(a-b)}$ and any of its cyclic permutations are all removable poles. We have tested this up to  large values of $p$ and seen that this cancelation of poles always happens so we conjecture this always happens for all values of $p$, but we have no proof. We will here obtain a general formula for $P_{i,reduced}$, which again will be a conjecture. For notational simplicity, we put
\bea
u &=& e^{\beta a}\cr
v &=& e^{\beta b}\cr
w &=& e^{\beta c}
\eea
which are subject to the constraint $u v w = 1$, and we put
\bea
x &=& e^{\beta}
\eea
In this notation, we have 
\bea
f_{hyper}(p) &=& \frac{1/w + v}{(1 - v/u) (1 - u/w) (1 - x^p u^p)} + \cycl\cr
f_{tensor}(p) &=& -\frac{1+v/w}{(1-v/u)(1 - u/w) (1 - x^p u^p)} + \cycl\cr
\eea
We rewrite these in the form
\bea
f_i &=& \frac{g_i}{1 - x^p u^p} + \cycl
\eea
where
\bea
g_{hyper} &=& -\frac{u+1}{(u-v)(u-w)}\cr
g_{tensor} &=& \frac{u(v+w)}{(u-v)(u-w)}
\eea
By adding the three cyclic terms, we get
\bea
f_i &=& \frac{a_i - x^p b_i + x^{2p} c_i}{(1 - x^p u^p)(1 - x^p v^p)(1 - x^p w^p)}
\eea
where
\bea
a_i &=& g_i+\cycl\cr
b_i &=& \(v^p+w^p\)g_i + \cycl\cr
c_i &=& \frac{1}{u^p}g_i + \cycl
\eea
Explicity we find
\bea
a_{hyper} &=& 0\cr
a_{tensor} &=& -1
\eea
For the other two terms, we have a rather complicated dependence on $p$. For the first few values of $p$ we find that 
\bea
b_{hyper}(1) &=& 1\cr
b_{hyper}(2) &=& 1 + u + v + w\cr
b_{hyper}(3) &=& u + v + w + u^2  + v^2 + w^2 + u v   + u w + v w 
\eea
\bea
b_{tensor}(1) &=& -u - v - w\cr
b_{tensor}(2) &=& -u^2 - v^2 - w^2 - u v - u w - v w \cr
b_{tensor}(3) &=& -u^3 - v^3 - w^3 - u^2 v - u v^2  - u^2 w - v^2 w - u w^2 - 
 v w^2  - 2 u v w
\eea
\bea
c_{hyper}(1) &=& -1\cr
c_{hyper}(2) &=& -1-\frac{1}{u}-\frac{1}{v}-\frac{1}{w}\cr
c_{hyper}(3) &=& -\frac{1}{u}-\frac{1}{v}-\frac{1}{w}-\frac{1}{u^2}-\frac{1}{v^2}-\frac{1}{w^2}-\frac{1}{uv}-\frac{1}{vw}-\frac{1}{wu}
\eea
\bea
c_{tensor}(1) &=& 0\cr
c_{tensor}(2) &=& u+v+w\cr
c_{tensor}(3) &=& 2 + \frac{u}{v}+\frac{v}{u}+\frac{u}{w}+\frac{w}{u}+\frac{v}{w}+\frac{w}{v}
\eea
We have the relations
\bea
c_{hyper}(p,u,v,w) &=& - b_{hyper}(p,1/u,1/v,1/w)\cr
c_{tensor}(p,u,v,w) &=& - b_{tensor}(p,1/u,1/v,1/w) -\frac{1}{u^p}-\frac{1}{v^p}-\frac{1}{w^p}
\eea
These relations determine $c_i$ once we know $b_i$. We also have the relation
\bea
b_{tensor}(p)+b_{hyper}(p+1)-b_{hyper}(p)+b_{hyper}(p-1) &=& 0
\eea
that we can use to determine $b_{tensor}$ once we know $b_{hyper}$. Our task has been reduced to determine $b_{hyper}$, before we have used the relation $uvw=1$. Let us now switch to a short notation. If $b_{hyper}(2) = 1+u+v+w$, then we will write this as $p=2$:$(100),(010),(001),(000)$ where $(100)$ represents the term $u$ and so on. We will also suppress all terms that are obtained by trivial permutations, so instead of writing out $(100),(010),(001)$, we will just write $(100)$. This way we get for the first few values of $p$ the following results,\footnote{We carried out this computation up to $p=6$ by using Mathematica.}
\bea
p=1 &:& [(000)]\cr
p=2 &:& [(100)],[(000)]\cr
p=3 &:& [(200),(110)],[(100)]\cr
p=4 &:& [(300),(210),(111)],[(200),(110)]\cr
p=5 &:& [(400),(310),(220),(211)],[(300),(210),(111)]\cr
p=6 &:& [(500),(410),(320),(221),(311)],[(400),(310),(220),(211)]
\eea
where we have grouped the elements into two classes. From this, we see the following  pattern
\bea
p=p &:& [(p-1,0,0),(p-2,1,0)...],[(p-2),(p-3,1,0)...]
\eea
where the first class of elements are all those elements whose entries sum up to $p-1$, and the second class are all those elements whose entries sum up to $p-2$. We have now in principle completed the computation, although the result has not been presented in an explicit way. This situation can be improved by restricting to $(u,v,w)=(u,1/u,1)$ where we get
\bea
b_{hyper}(p,u,1/u,1) &=& (p-1) \(u+u^{-1}\) + (p-2) \(u^2+u^{-2}\) + \cdots + \(u^{p-1}+u^{-(p-1)}\) + p\cr
c_{hyper}(p,u,1/u,1) &=& -b_{hyper}(p,u,1/u,1)
\eea
\bea
b_{tensor}(p,u,1/u,1) &=& -(p-1) \(u+u^{-1}\) - (p-2) \(u^2+u^{-2}\) - \cdots - \(u^{p-1}+u^{-(p-1)}\) - p\cr
&& - \(u^p + u^{-p}\)\cr
c_{tensor}(p,u,1/u,1) &=& (p-1) \(u+u^{-1}\) + (p-2) \(u^2+u^{-2}\) + \cdots + \(u^{p-1}+u^{-(p-1)}\)+p\cr
&& -1
\eea
For the sum we have a closed form, 
\ben
b_{hyper}(p,u,1/u,1) &=& \(\frac{u^{p/2}-u^{-p/2}}{u^{1/2}-u^{-1/2}}\)^2\label{nice}
\een
Using this result, we get
\bea
f_{hyper}(p,u,1/u,1) &=& -\(\frac{u^{p/2}-u^{-p/2}}{u^{1/2}-u^{-1/2}}\)^2 \frac{x^p+x^{2p}} {(1 - x^p u^p)(1 - x^p v^p)(1 - x^p w^p)}\Bigg|_{v=1/u,w=1}\cr
f_{tensor}(p,u,1/u,1) &=& \(\frac{u^{p/2}-u^{-p/2}}{u^{1/2}-u^{-1/2}}\)^2 \frac{x^p+x^{2p}} {(1 - x^p u^p)(1 - x^p v^p)(1 - x^p w^p)}\Bigg|_{v=1/u,w=1}\cr
&& + \frac{-1+\(u^p+u^{-p}\)x^p - x^{2p}}{(1 - x^p u^p)(1 - x^p v^p)(1 - x^p w^p)}\Bigg|_{v=1/u,w=1}\cr
\eea
The second term in the second line can be simplified by noting that
\bea
(1 - x^p u^p)(1 - x^p u^{-p}) &=& 1 - x^p (u^p+u^{-p}) + x^{2p}
\eea
Then, by adding the two contributions, we get the result,
\bea
f_{M5}(p,u,1/u,1) &=& \frac{1}{-1 + x^p}
\eea
We notice that surprisingly this M5 brane generating function does not depend on the squashing parameter $a$. 

To get a nontrivial dependence on squashing parameters for the M5 brane generating function at $m_H = 1/2$, we shall consider generic $a,b,c$. Then we shall return to our result above. To streamline the notation, we define
\bea
s_q(u,v,w) &=& \sum_{r+s+t=q} u^r v^s w^t
\eea
and
\bea
Q_{q_1,q_2,...}(u,v,w) &=& s_{q_1}(u,v,w)+s_{q_2}(u,v,w)+...
\eea
Then we have  
\ben
b_{hyper}(p,u,v,w) &=& Q_{p-1,p-2}(u,v,w)\label{sums}
\een
Let us further define the set 
\bea
\Q_{q_1,q_2,...} &=& \{r,s,t|r+s+t=q_1\} \cup \{r,s,t|r+s+t=q_2\}\cup...
\eea
We then conjecture the following general expression,
\ben
f_{hyper}(p,\beta,a,b,c) &=& \sum_{r+s+t \in \Q_{p-1,p-2}} \frac{\cosh \[\frac{p\beta}{2} + \beta(ar + bs +c t)\]}{4\sinh\frac{p\beta}{2}(1+a)\sinh\frac{p\beta}{2}(1+b)\sinh\frac{p\beta}{2}(1+c)}\label{generalhyper}
\een
As consistency checks, we note that if we take the limit $u,v,w\rightarrow 1$, then the sum (\ref{sums}) reduces to 
\bea
b_{hyper}(p,1,1,1) = \frac{p(p+1)}{2} + \frac{(p-1)p}{2} = p^2
\eea
and if we put $(u,v,w)=(u,1/u,1)$ in (\ref{generalhyper}), then it reduces to (\ref{nice}). 

Let us move on to the tensor multiplet. We clearly seem to have 
\bea
b_{tensor}(p,u,v,w) &=& -Q_{p,p-3}(u,v,w)
\eea
As checks, we see that in special cases, this reduces to the previous results,
\bea
b_{tensor}(p,1,1,1) &=& -p^2 - 2\cr
b_{tensor}(p,u,1/u,1) &=& -\(\frac{u^{p/2}-u^{-p/2}}{u^{1/2}-u^{-1/2}}\)^2 - \(u^p+u^{-p}\)
\eea
We are then ready to conjecture the general result for the generating function,
\bea
f_{tensor}(p,x,u,v,w) &=& \frac{-1+x^p Q_{p,p-3} + x^{2p} \(Q^{\vee}_{p,p-3}-u^{-p}-v^{-p}\)}{(1-x^p u^p)(1-x^p v^p)(1-x^p w^p)}
\eea
where we define 
\bea
Q_{p,p-3}^{\vee}(\beta) &=& Q_{p,p-3}(-\beta)
\eea
We can take out a simple term from this and write the rest in a manifestly antisymmetric form,
\bea
f_{tensor} &=& - \frac{1}{1-x^p w^p}\cr
&& + \frac{x^{\frac{p}{2}} \(Q_{p,p-3}^{\vee}-u^{-p}-v^{-p}\)+x^{-\frac{p}{2}}\(Q_{p,p-3} - u^p -v^p\)}{\((xu)^{\frac{p}{2}}-(xu)^{-\frac{p}{2}}\) \((xv)^{\frac{p}{2}}-(xv)^{-\frac{p}{2}}\) \((xw)^{\frac{p}{2}}-(xw)^{-\frac{p}{2}}\)}
\eea
which can also be written as
\ben
f_{tensor} &=& \frac{e^{-p\beta(1+c)}}{1-e^{-p\beta(1+c)}}\cr
&& + \frac{\cosh \frac{p\beta(1-2a)}{2} + \cosh \frac{p\beta(1-2b)}{2} - \sum_{r,s,t\in \Q_{p,p-3}} \cosh \frac{p\beta\(1-\frac{2}{p}(ra+sb+tc)\)}{2}}{4\sinh\frac{p\beta(1+a)}{2} \sinh\frac{p\beta(1+b)}{2} \sinh\frac{p\beta(1+c)}{2}}\label{generaltensor}
\een
It is easy to see that we reproduce the previously known result if we put $p=1$. For $p=1$ we have
\bea
&&\sum_{r,s,t\in \Q_{p,p-3}} \cosh \frac{p\beta}{2}\(1-\frac{2}{p}(ra+sb+tc)\)\Bigg|_{p=1} \cr
&=& \cosh \frac{\beta(1-2a)}{2} + \cosh \frac{\beta(1-2b)}{2} + \cosh \frac{\beta(1-2c)}{2}
\eea
and so we get
\ben
f_{tensor} &=& \frac{e^{-\beta(1+c)}}{1-e^{-\beta(1+c)}}
 - \frac{\cosh \frac{\beta(1-2c)}{2}}{4\sinh\frac{\beta(1+a)}{2} \sinh\frac{\beta(1+b)}{2} \sinh\frac{\beta(1+c)}{2}}\label{tensorsplit1}
\een

Let us now return to the hypermultiplet. To understand how to generalize to general $m_H$, it is sufficient to just look at say the case with $p=2$ for which we get
\bea
f_{hyper}(2,x,u,v,w,t) &=& \frac{(t^{-1}+t(u+v+w))x^{-1} + (t+t^{-1}(u^{-1}+v^{-1}+w^{-1}))x}{(ux-u^{-1}x^{-1})(vx-v^{-1}x^{-1})(wx-w^{-1}x^{-1})}
\eea
If we express this in the form
\bea
f_{hyper}(2,x,u,v,w,t) &=& \sum_{r,s,t\in \Q_0} \frac{t^{-1} u^r v^s w^t + t u^{-r} v^{-s} w^{-t}}{(ux-u^{-1}x^{-1})(vx-v^{-1}x^{-1})(wx-w^{-1}x^{-1})} \cr
&& + \sum_{r,s,t\in \Q_1} \frac{t^{-1} u^r v^s w^t + t u^{-r} v^{-s} w^{-t}}{(ux-u^{-1}x^{-1})(vx-v^{-1}x^{-1})(wx-w^{-1}x^{-1})}
\eea
then it seems clear that this should generalize as 
\ben
f_{hyper}(p,\beta,a,b,c,\epsilon) &=& \sum_{r,s,t\in \Q_{p-2}} \frac{\cosh \[\frac{p\beta}{2} - \beta(ar + bs +c t) - \beta \epsilon\]}{4\sinh\frac{p\beta}{2}(1+a)\sinh\frac{p\beta}{2}(1+b)\sinh\frac{p\beta}{2}(1+c)}\cr
&+&\sum_{r,s,t\in \Q_{p-1}} \frac{\cosh \[\frac{p\beta}{2} - \beta(ar + bs +c t) + \beta \epsilon\]}{4\sinh\frac{p\beta}{2}(1+a)\sinh\frac{p\beta}{2}(1+b)\sinh\frac{p\beta}{2}(1+c)}\label{hyperconj}
\een
As two consistency checks for our conjectured formula (\ref{hyperconj}), we first notice that the small $\beta$ expansion agrees with (\ref{hypexpand}) for arbitrary $p$, and second, if we put $a=b=c=0$ and keep $p$ arbitrary, then we reproduce (\ref{hypprevious}) to all orders in $\beta$.

\subsection{Dualization using the plethystic method}
Having obtained the generating functions in their fully reduced forms, we are now ready to dualize these generating functions using the plethystic method as we outline in the appendix following \cite{Kim:2012qf}. We have already extracted the Stefan-Boltzmann terms. Let us move on to compute the integrals
\bea
\frac{1}{2} \int_{-\infty}^{\infty} \frac{ds}{s} f_{i,reg}(s)
\eea
for $i$ running over hyper and tensor multiplets. If there are only simple poles, the instanton contribution is computed in a similar way,
\bea
\sum_{k=1}^{\infty} \int_{-\infty}^{\infty} \frac{ds}{s} f_{i,reg}(s) e^{2\pi i k s/\beta}
\eea
and if simple poles are located at (\ref{simplepoles}), then the sum over $k$ becomes a geometric series,
\bea
\sum_{k=1}^{\infty} e^{-\frac{4\pi^2 kn}{p\beta(1+a)}} &=& \frac{e^{-\frac{4\pi^2 n}{p\beta(1+a)}}}{1-e^{-\frac{4\pi^2 n}{p\beta(1+a)}}}
\eea

\subsubsection{The hypermultiplet}
We note that this integral does not depend on $\beta$ so it should be part of the perturbative contibution from the 5d viewpoint. The advantage with turning on generic $a,b,c$ parameters is that there now will appear only contributions from simple poles located at
\ben
s &=& \frac{2\pi i n}{p(1+a)}\cr
s &=& \frac{2\pi i n}{p(1+b)}\cr
s &=& \frac{2\pi i n}{p(1+c)}\label{simplepoles}
\een
and by closing the contour in the upper halfplane, we will pick up contributions only from those poles with $n=1,2,3,...$. We expand around a pole,
\bea
\sinh\frac{p s (1+a)}{2} &=& (-1)^n \frac{p(1+a)}{2} \(s-\frac{2\pi i n}{p(1+a)}\) + ...
\eea
and get the corresponding residue 
\bea
H_n(a,b,c) &=& -\sum_{\Q_{p-2}} \frac{1}{2n} \frac{\cos \frac{\pi n}{1+a}\(2+a-\frac{2}{p}(a r+bs + ct) - \frac{2\epsilon}{p}\)}{\sin \frac{\pi n(b-a)}{1+a} \sin \frac{\pi n(c-a)}{1+a}}\cr
&& -\sum_{\Q_{p-1}} \frac{1}{2n} \frac{\cos \frac{\pi n}{1+a}\(2+a-\frac{2}{p}(a r+bs + ct) + \frac{2\epsilon}{p}\)}{\sin \frac{\pi n(b-a)}{1+a} \sin \frac{\pi n(c-a)}{1+a}}
\eea
after some computation. The contribution from the other two poles can be obtained by cyclic permutations of $a,b,c$. Thus we have obtained
\bea
\frac{1}{2} \int_{-\infty}^{\infty} \frac{ds}{s} f_{hyper,reg}(s) &=& \sum_{n=1}^{\infty} \(H_n(a,b,c)+H_n(b,c,a)+H_n(c,a,b)\)
\eea
If we take $p=1$ we get only the contribution from $r=s=t=0$ from the sum in the second line,
\bea
H_n(a,b,c) &=& -\frac{1}{2n} \frac{\cos \frac{\pi n}{1+a}\(2+a + 2\epsilon\)}{\sin \frac{\pi n(b-a)}{1+a} \sin \frac{\pi n(c-a)}{1+a}}
\eea
which then leads to a result that is in a good agreement with eq (2.67) in \cite{Kim:2012qf}. 

To take the unsquashed limit, we first assume that $\epsilon = 0$ and define 
\bea
{\t H}_n(\lambda,a,b,c) &=& H_n(\lambda a,\lambda b,\lambda c) + H_n(\lambda b,\lambda c,\lambda a) + H_n(\lambda c,\lambda a,\lambda b)
\eea
and then Taylor expand 
\bea
{\t H}_n(\lambda,a,b,c) &=& -\frac{1}{2 n^3 \pi^2} + \frac{2 abc}{15} n \pi^2 \lambda^3 + \O(\lambda^4)
\eea
and we get
\bea
\frac{1}{2}\sum_{n=1}^{\infty} \(H_n(a,b,c) + H_n(b,c,a) + H_n(c,a,b)\) &=& -\frac{p^2}{4\pi^2} \zeta(3) + \frac{p^2}{15} a b c \zeta(-1) + \O(|a,b,c|^4)
\eea
where the factor $p^2$ comes from the sum $\sum_{\Q_{p-1,p-2}} 1 = p^2$. This result shows that the unsquashed limit is smooth and we reproduce our previous result (\ref{hyperpert}), if we interpret the sum by means of zeta function regularization. 

If we keep $\epsilon$ nonzero, then there will be a correction to this result on the form
\bea
{\t H}_n(\lambda,a,b,c) &=& -\frac{1}{2 n^3 \pi^2} + \O(\epsilon^3,\lambda^4)
\eea
which is still consistent with our previous result that we computed up to order $\O(\epsilon^2)$.

Let us move on to the instanton contribution. We then consider the sum
\bea
\sum_{n=1}^{\infty} \sum_{k=1}^{\infty} \(H_{n,k}(a,b,c) + H_{n,k}(b,c,a) + H_{n,k}(c,a,b)\)
\eea
where we define 
\bea
H_{n,k}(a,b,c) &=& H_n(a,b,c) e^{-\frac{4\pi^2 n k}{p\beta(1+a)}}
\eea
To take the unsquashed limit, we define 
\bea
{\t K}_{n,k}(\lambda;a,b,c) = K_{n,k}(\lambda a,\lambda b,\lambda c) + K_{n,k}(\lambda b,\lambda c,\lambda a) + K_{n,k}(\lambda c,\lambda a,\lambda b)
\eea
and expand $\t K$ up zeroth order in $\lambda$,
\bea
{\t K}_{n,k}(\lambda;a,b,c) &=& \(\frac{2k}{\beta n^2 p} + \frac{1}{2\pi^3 n^3} + \frac{4\pi^2 k^2}{p^2 \beta^2 n}\) e^{-\frac{4 \pi^2 k n}{p\beta}} + \O(\lambda)
\eea
We notice that no singular terms appear in this expansion, and that the finite term is independent of $a,b,c$, which means that the unsquashed limit is well-defined and does not depend on how we let $a,b,c$ approach to zero as long as $a+b+c =0$. This result then leads to a perfect agreement with our previous result in eq (\ref{hyperinst}). Hence also for the instanton contribution, the unsquashed limit is smooth.

\subsubsection{The tensor multiplet}
For the tensor multiplet, the cyclic symmetry in $a,b,c$ is hidden once we separate out the first term in (\ref{generaltensor}). We will now dualize the terms in the second line in (\ref{generaltensor}) which are antisymmetric under $\beta\rightarrow -\beta$ so that we can apply the plethystic dualization method on these terms alone.\footnote{Later we will take back this statement, due to regularization issues.} We get the following residues at a given $n$ when we compute the integral $\int ds f_{tensor,reg}(s)/s$ from the first two terms on the second line of eq (\ref{generaltensor}),
\ben
&&-\frac{\cos\frac{\pi n 3a}{1+a} + \cos\frac{\pi n(a+2b)}{1+a}}{2n\sin\frac{\pi n(b-a)}{1+a} \sin\frac{\pi n(c-a)}{1+a}}\cr
&&-\frac{\cos\frac{\pi n(b+2a)}{1+b} + \cos\frac{\pi n3b}{1+b}}{2n\sin\frac{\pi n(a-b)}{1+b}\sin\frac{\pi n(c-b)}{1+b}}\cr
&&-\frac{\cos\frac{\pi n(c+2a)}{1+c} + \cos\frac{\pi n (c+2b)}{1+c}}{2n\sin\frac{\pi n(a-c)}{1+c}\sin\frac{\pi n(b-c)}{1+c}}\label{three}
\een
These are cyclic permutations up to the following terms
\bea
-\frac{\cos \frac{\pi n 3b}{1+b} - \cos \frac{\pi n(b+2c)}{1+b}}{2n\sin\frac{\pi n(a-b)}{1+b}\sin\frac{\pi n(c-b)}{1+b}} &=& \frac{1}{n}
\eea
and
\bea
-\frac{\cos \frac{\pi n 3a}{1+a} - \cos \frac{\pi n(a+2c)}{1+a}}{2n\sin\frac{\pi n(b-a)}{1+a} \sin\frac{\pi n(c-a)}{1+a}} &=& \frac{1}{n}
\eea
where we have used the trigonometric identity $2 \sin A \sin B = \cos(A-B)-\cos(A+B)$. The last term on the second line in eq (\ref{generaltensor}) contributes something that has already cyclic permutation symmetry,
\bea
\sum \frac{\cos\frac{\pi n(a+2(ra+sb+tc)/p)}{1+a}}{2n\sin\frac{\pi n(b-a)}{1+a} \sin\frac{\pi n(c-a)}{1+a}} + \cycl
\eea
Summing all the contributions coming from the second line in eq (\ref{generaltensor}) we get the result
\bea
\frac{1}{2} \int_{-\infty}^{\infty} \frac{ds}{s} f_{tensor,reg,2^{nd}\quad line}(s)
 &=& \sum_{n=1}^{\infty} \frac{1}{2n} \(t_n(a,b,c)+t_n(b,c,a)+t_n(c,a,b) + 2\)
\eea
where
\bea
t_n(a,b,c) &=& -\frac{\cos\frac{\pi n (b-c)}{1+a}}{\sin\frac{\pi n(b-a)}{1+a} \sin\frac{\pi n(c-a)}{1+a}} +\sum_{r,s,t\in Q_{p,p-3}} \frac{\cos\frac{\pi n(a+2(ra+sb+tc)/p)}{1+a}}{2\sin\frac{\pi n(b-a)}{1+a} \sin\frac{\pi n(c-a)}{1+a}}
\eea
For $p=1$ this reduces to
\bea
t_n(a,b,c) &=& \frac{1}{2}T_n(a,b,c)-1
\eea
where
\bea
T_n(a,b,c) &=& \frac{\cos\frac{\pi n(b-c)}{1+a}}{2 \sin\frac{\pi n(b-a)}{1+a} \sin\frac{\pi n(c-a)}{1+a}}
\eea
and we get
\ben
\frac{1}{2} \int_{-\infty}^{\infty} \frac{ds}{s} f_{tensor,reg}(s)
 &=& \sum_{n=1}^{\infty} \frac{1}{2n} \(T_n(a,b,c)+T_n(b,c,a)+T_n(c,a,b) - 1\)\label{seek}
\een
This is in good agreement with eq (2.66) in \cite{Kim:2012qf}. Let us expand around $a=b=c=0$ up to cubic order. For simplicity let us take $p=1$. Thus we define
\bea
{\t T}_n(\lambda,a,b,c) &=& T_n(\lambda a,\lambda b,\lambda c)+T_n(\lambda b,\lambda c,\lambda a)+T_n(\lambda c,\lambda a,\lambda b)
\eea
The small $\lambda$-expansion reads
\bea
{\t T}_n(\lambda,a,b,c) &=& \frac{1}{4\pi^3} \frac{1}{n^3} + \frac{1}{2n} - \frac{3\pi^2}{5} abc n \lambda^3 + \O(\lambda^4)
\eea
The problematic term $1/(2n)$ cancels against $-1/(2n)$. Then after carrying out the summation over $n$ using zeta function regularization, we get
\bea
\frac{1}{2} \int_{-\infty}^{\infty} \frac{ds}{s} f_{tensor,reg}(s) &=& \frac{1}{4\pi^3} \zeta(3) - \frac{3\pi^2}{5} abc \zeta(-1) + \O(|a,b,c|^4)
\eea
Up to cubic order, this expression is completely symmetric in $a,b,c$ and there is no reason to expect this symmetry will be broken at higher orders. But there is obviosly a problem here since if we dualize the first simple term in (\ref{tensorsplit1}), then we get a term proportional to
\ben
\frac{1}{2}\log \(\beta(1+c)\)\label{first}
\een
which breaks the permutation symmetry among $a,b,c$. To restore it by adding the contribution that we get from dualizing the second term, the second term can not lead to a result that is completely symmetric in $a,b,c$ \cite{Kim:2012qf}. We may not be allowed to dualize the second term by the plethystic method, since we can not dualize the first simple term in (\ref{tensorsplit1}) by the plethystic method. We demonstrate this fact at the end of appendix C. To use the plethystic method to dualize the first term, we need to regularize a divergent plethystic sum $\sum 1/n$ which can not be regularized using the zeta function. We may for instance multiply the generating function by a gaussian regulator $e^{-\epsilon \beta^2}$ (which is symmetric in $\beta$) and then at the end take $\epsilon \rightarrow 0$. This effectively places a cutoff at $n \sim N_{\epsilon} \sim 1/\sqrt{\epsilon}$ for the sum over $n$. The same regularization should then be used throughout, hence to the whole expression in (\ref{tensorsplit1}). Now, if we apply this regularization to the second term as well, it will amount to a multiplication of each of the three terms at the three lines in eq (\ref{three}) by their corresponding regulator factor $e^{-\epsilon s^2}$ where $s$ is evaluated at the three poles in (\ref{simplepoles}) for each line respectively. The correction is proportional to $\epsilon$ for the zeroth order term in $\lambda$, and so it goes to zero as we take $\epsilon$ to zero. But interesting regularization effects can show up at cubic order where we are regularizing a divergent sum $\sum n$. This will then contain a divergent piece $\sum_{n=1}^{N_{\epsilon}} n \sim N_{\epsilon}^2 \sim 1/\epsilon$, which is canceled against the order $\epsilon$ correction to (\ref{three}), which is asymmetric in $a,b,c$. A detailed such computation would involve the error function and so it would be quite involved. Let us here content ourselves with noting that in the limit $a,b,c\rightarrow 0$ we reproduce the correct result, which agrees with our previous results that we obtained by rigorous methods. When $a,b,c$ is away from zero, we may not have got the entirely correct result by the plethystic dualization method, but the error should be well-confined and small as long as $\lambda$ is small. The error we made is only concerning the perturbative part. No instanton sum is affected by this issue, since the instanton sums are convergent.

Let us move on to the instanton sum. We define
\bea
T_{n,k}(a,b,c) &=& T_n(a,b,c) e^{-\frac{4\pi^2 n k}{\beta(1+a)}}
\eea
and 
\bea
{\t T}_{n,k}(\lambda,a,b,c) &=& T_{n,k}(\lambda a,\lambda b,\lambda c)+T_{n,k}(\lambda b,\lambda c,\lambda a)+T_{n,k}(\lambda c,\lambda a,\lambda b)
\eea
This has the small $\lambda$-expansion
\bea
{\t T}_{n,k}(\lambda,a,b,c) &=& \(\frac{k}{\beta n^2} + \frac{1}{2n} + \frac{1}{4\pi^2 n^3} + \frac{2 \pi^2 k}{\beta^2 n}\) e^{-\frac{4\pi^2 k n}{\beta}} + \O(\lambda^2)
\eea
By finally adding the contribution coming from the first term in (\ref{generaltensor}), which gives the contribution
\bea
\frac{\pi^2}{6\beta} + \frac{1}{2} \log\frac{\beta}{2\pi}
\eea
we reduce to our previous result (\ref{unsquashedtensor}) in the unsquashed limit.\footnote{For the comparison with (\ref{unsquashedtensor}) we should remember to multiply $\t T_k$ by $2$ since we defined this out of $T$ that is in the perturbative part where we have the factor of $1/2$ multiplying the integral $\int ds f_{reg}/s$, while there is no such factor $1/2$ for the corresponding integral for the instanton contributions.}

\section{Asymptotic S-duality}\label{sectS}
In \cite{Shaghoulian:2016gol} it was argued that for the geometry $S^1_{\beta} \times (S^5_r/\mb{Z}_p)$ there appears an emergent rectangular $T^2$ spanned by $S^1_{\beta}$ and the Hopf fiber of $S^5$, in the limit when $p$ becomes very large. It was then argued that there would be an S-duality associated with this emergent $T^2$. The radius of the temporal $S^1_{\beta}$ is $\beta$, while the radius of the Hopf fiber is $2\pi r/p$ where $r$ is the radius of $S^5$. In order to exhange these two circles, it is convenient to follow \cite{Shaghoulian:2016gol} and put $
\beta = 2\pi r/k$ for some integer $k$. The S-dual geometry will then correspond to $S^1_{\beta_D} \times (S^5_r/{\mb{Z}_k})$ with $\beta_D = 2\pi r/p$. For a $T^2$ to emerge on both sides of the duality, we need to assume that both $p$ and $k$ are very large integer numbers. For the duality to relate high and low temperature behaviors, we need to in addition assume that $k<<p$. Then $\beta_D << \beta$ which means that the S-dual geometry corresponds to the high temperature side of the duality.

We will now test whether asymptotic S-duality holds, which we can do since we know both the low temperature and the high temperature behaviors of the logarithm our indices (free energies). The high temperature behavior of the free energy is governed by the Stefan-Boltzmann terms, in which we shall put $\beta = 2\pi r/k$. The low temperature behavior is governed by the Casimir energy computed on $S^5/{\mb{Z}_k}$ multiplied by $\beta_D = 2\pi r/p$. 

\subsection{Accidental asymptotic S-duality for the index}
We begin with listing our results for the Stefan-Boltzmann terms
\bea
\beta F_{tensor} &=& \frac{\pi^4}{45 Np \beta^3}-\frac{(6+5(ab+bc+ca))\pi^2}{36Np\beta}\cr
\beta F_{hyper} &=& -\frac{\pi^4}{45 Np\beta^3} - \frac{(ab+bc+ca+6(\epsilon+\epsilon^2))\pi^2}{36 N p\beta}
\eea
and for the Casimir energies
\bea
E_{tensor} &=& -\frac{p^3}{240} - \frac{p}{24} + \frac{a^4+b^4+c^4-36abc}{480 N p}\cr
E_{hyper} &=& \frac{p^3}{240} + \frac{p\epsilon}{24} - \frac{\epsilon^3}{12 N p} - \frac{\epsilon^4}{24 N p}\cr
&& - \frac{a^4+b^4+c^4+4abc}{480 N p} \cr
&& - \frac{2(ab+bc+ca)+abc}{24Np}\epsilon\cr
&& -\frac{ab+bc+ca}{24Np}\epsilon^2
\eea
We are now ready to test asymptotic S-duality. In the Stefan-Boltzmann terms we put $\beta = 2\pi r/k$ and get
\bea
\beta F_{tensor} &=& \frac{\pi k^3}{360 p} - \frac{(6+5(ab+bc+ca)\pi k}{72Np}\cr
\beta F_{hyper} &=& -\frac{\pi k^3}{360 p} - \frac{(ab+bc+ca)\pi k}{72Np}
\eea
and for the Casimir energies we replace $p$ by $k$ and multiply by $\beta_D = 2\pi r/p$ to get
\bea
\beta_D E_{tensor} &=& -\frac{\pi k^3}{120 p} - \frac{\pi k}{12 p} + \frac{\pi\(a^4+b^4+c^4-36abc\)}{240 N k p}\cr
\beta_D E_{hyper} &=& \frac{\pi k^3}{120p} + \frac{\pi k\epsilon}{12p} - \frac{\pi \epsilon^3}{6 N kp} - \frac{\pi \epsilon^4}{12 N k p}\cr
&& - \frac{\pi\(a^4+b^4+c^4+4abc\)}{240 N k p} \cr
&& - \frac{\pi\(2(ab+bc+ca)+abc\)}{12Nk p}\epsilon\cr
&& -\frac{\pi\(ab+bc+ca\)}{12Nkp}\epsilon^2
\eea
S-duality for $(1,0)$ supermultiplets would hold if we had $\beta F_i = \beta_D E_i$ for $i=\{tensor,hyper\}$. Clearly we do not have such an S-duality. Things improve if we consider $(2,0)$ theory for which we put $\epsilon =0$. Then we have for the sum
\bea
\beta F_{M5} &=& - \frac{\pi k}{12 N p} - \frac{\(ab+bc+ca\)\pi k}{12 N p}\cr
&=& -\frac{\pi k}{12 p} + \frac{abc \pi k}{12 Np}
\eea
where in the second step we used $N = 1 + ab + bc + ca + abc$. We also have 
\bea
\beta_D E_{M5} &=& - \frac{\pi k}{12 p} - \frac{abc}{6 N k p}
\eea
Thus we have $\beta F_{M5} = \beta_D E_{M5}$ if and only if $abc=0$. Moreover, if we put $a=b=c=0$ and keep $\epsilon$ arbitrary, then we have
\bea
\beta F_{M5} &=& - \frac{\pi k}{12p} - \epsilon \frac{\pi k}{12 p} - \epsilon^2 \frac{\pi k}{12 p}\cr
\beta_D E_{M5} &=& - \frac{\pi k}{12 p} + \epsilon \frac{\pi k}{12 p} - \epsilon^3 \frac{\pi}{6 k p} -\epsilon^4 \frac{\pi}{12 k p}
\eea
Hence only when $\epsilon = 0$ and $abc = 0$ can we have asymptotic S-duality. We believe that this asymptotic S-duality that we see here is rather accidental, and a result of two competing effects. On the one hand we have an increased amount of supersymmetry at $\epsilon =0$ and $a=b=c=0$. On the other hand we have with increased amount of supersymmetry also further cancellation of leading powers that lowers the leading power from $T^3$ down to $T$ in the Stefan-Boltzmann terms in the large $T$ limit, where $T = 1/\beta$ is the temperature. The cancelation of leading power appears to make asymptotic S-duality less likely to hold, but then increased supersymmetry apparently compensates for that so that we can see asymptotic S-duality nevertheless.

But asymptotic S-duality was expected to hold by a very general argument in \cite{Shaghoulian:2016gol}, and thus we would not expect to only see this duality by some accident. We have found that in general there is no such asymptotic S-duality for indices, other than for a rather accidental choice of parameters. We could then ask ourselves why this is so. We believe that the answer is due to the fact that in 5d the generic leading term in the Stefan-Boltzmann law should generically grow like $T^5$ for large $T$ and asymptotic S-duality is expected only for this leading term. But for the supersymmetric indices, we have no such high power leading term as $T^5$ due to supersymmetric cancelation. The asymptotic S-duality only holds in the very large $p$ and $k$ limits and thus is expected to be seen only for the $T^5$ term in the Stefan-Boltzmann law, which grows like $k^5$ when we put $\beta \sim 1/k$. To see those terms, we may instead consider the contribution to the index coming from each individual field before the cancelation has taken place. Or we may consider the partition function rather than the index. Indeed, here we will see asymptotic S-duality that seems to be generic, rather than accidental.

\subsection{Generic asymptotic S-duality for individual fields}
We will now demonstrate asymptotic S duality at leading order $T^5$ for each individual field in the $(1,0)$ tensor multiplet.

For the fields in the tensor multiplet, the scalar field (S), the tensor gauge field (T) and the Weyl fermions (F), we have on a round $S^5$ the following refined degeneracies \cite{Bak:2016vpi}
\bea
d^S_n(\omega) &=& \sum_{m=0}^n d_{m,n-m} \omega^{2m-n}\cr
d^T_n(\omega) &=& \sum_{m=0}^{n-1} \(d_{m,n - m - 1} \omega^{2 m - n - 2} + d_{m, n - m} \omega^{2 m - n} + d_{m, n - m + 1} \omega^{2 m - n + 2}\)\cr
d^F_n(\omega) &=& \sum_{m=0}^n \(d_{m,n-m}\omega^{2m-n-3/2}+d_{m,n-m+1}\omega^{2m-n+1/2}\)
\eea
where
\bea
d_{p,q} &=& \frac{1}{2}(p + 1) (q + 1) (p + q + 2)
\eea
Although we have the relation $4d^S_n = d^F_{n-1}+d^F_n$ for the unrefined degeneracies, this relation does not extend to the refined case. 

We define
\bea
D_n^S(\omega) &=& d_{n-2}^S(\omega)\cr
D_n^T(\omega) &=& d_{n-2}^T(\omega)\cr
D_n^F(\omega) &=& d_{n-3}^F(\omega) \omega^{-3/2} + d_{n-2}^F(\omega) \omega^{3/2}
\eea
and then we have the refined generating functions
\bea
f^i(\beta,\omega) &=& \sum_{n=0}^{\infty} D_n^i(\omega) e^{-\beta n}
\eea
for $i=S,T,F$ and where one may check that we can extend the sum all the way down to $n=0$ since for $n=0,1$ there is no nonzero contribution to the sum. Although there are no nice and simple explicit expressions for these refined dimensions, we are able to repackage these refined dimensions into manageable closed form expressions for the refined generating functions,
\bea
f^S(\beta,\omega) &=& \frac{e^{-2 \beta}-e^{-4 \beta}}{(1-\omega e^{-\beta})^3 (1 - \omega^{-1} e^{-\beta})^3}\cr
f^T(\beta,\omega) &=& \frac{e^{-3 \beta}(\omega^{-3} + 3 \omega^{-1} + 6 \omega)
-  3 e^{-4 \beta}(\omega^{-2} +\omega^2+3) + 3 e^{-5 \beta} (\omega^{-1}  + \omega) - e^{-6 \beta}}{(1-\omega e^{-\beta})^3 (1 - \omega^{-1} e^{-\beta})^3} \cr
f^F(\beta,\omega) &=& \frac{e^{-2 \beta}(1+ 3 \omega^2)+ e^{-3 \beta}(\omega^{-3} + 3 \omega^{-1} - 3 \omega  - \omega^3)
-e^{-4 \beta}(1+3 \omega^{-2})}{(1-\omega e^{-\beta})^3 (1 - \omega^{-1} e^{-\beta})^3}
\eea
We get a simplification when we compute the refined tensor multiplet generating function,
\bea
f_{tensor} = f^S+f^T-f^F = \frac{e^{-3 \beta} \omega^3-3 e^{-2 \beta} \omega^2}{(1 - \omega e^{-\beta})^3}
\eea
but we will not consider this object here, but rather the contributions from the individual fields. To obtain the corresponding generating functions on lens space $L(p;1,1)$, we expand the denominator in an infinite series
\bea
f^i(\beta,\omega) &=& \frac{\sum_{\lambda} \omega^{\lambda} f^i_{\lambda}(\beta)}{(1-\omega e^{-\beta})^3(1-\omega^{-1} e^{-\beta})^3}\cr
&=& \sum_{\lambda} f^i_{\lambda}(\beta) \sum_{n,m=0}^{\infty} \frac{1}{4} (n+1)(n+2)(m+1)(m+2) e^{-\beta(n+m)} \omega^{n-m+\lambda}
\eea
and then we sum over $\omega$ running over all the $p$ distinct $p$-th roots of unity that will put $n-m + \lambda = p q$. Let us assume that $p$ is sufficiently large, such that
\bea
\lambda < p
\eea
Let us furthermore restrict ourselves to the case that $\lambda\geq 0$. Since $n=m+pq-\lambda\geq 0$, we see that for $q\geq 1$ there will be no further restriction on $m$ coming from requiring that $m\geq \lambda - pq$ since by our assumptions we will have $\lambda - pq<0$. Hence part of our sum will consist of 
\bea
\sum_{q=1}^{\infty} \sum_{m=0}^{\infty} f(m+pq-\lambda,m)
\eea
Let us next assume that $q\leq -1$. We then bring this into $q\geq 1$ by first exchanging $m$ and $n$ assuming that the summand has this exchange symmetry, and next replacing $\lambda$ by $\lambda' = -\lambda$ which is negative. We then need to analyse the case when $-p<\lambda'\leq 0$. Here we find no restrictions at all, so we have the contribution
\bea
\sum_{q=1}^{\infty} \sum_{m=0}^{\infty} f(m+pq+\lambda,m)
\eea
Now only remains the case when $q=0$. Then $n=m-\lambda$. If $\lambda\geq 0$, then we have the contribution
\bea
\sum_{n=0}^{\infty} f(n,n+\lambda) &=& \sum_{m=0}^{\infty} f(m+\lambda,m)
\eea
If $\lambda$ is negative, we have the contribution
\bea
\sum_{m=0}^{\infty} f(m-\lambda,m)
\eea
Summing all contributions, we get a quantity that we call $S_{\lambda}$,
\bea
S_{\lambda} &=& \sum_{q=1}^{\infty} \sum_{m=0}^{\infty} \(f(m+pq-\lambda,m)+f(m+pq+\lambda,m)\)+\sum_{m=0}^{\infty} f(m+|\lambda|,m)
\eea
where we shall take 
\bea
f(n,m) &=& \frac{1}{4}(n+1)(n+2)(m+1)(m+2) e^{-\beta(n+m)}
\eea
We then get the lensed indices as follows,
\bea
f^S &=& (e^{-2\beta} - e^{-4\beta}) S_0\cr
f^T &=& e^{-3\beta} (S_3 + 9 S_1) - e^{-4\beta} 3(2S_2 +3 S_0) + e^{-5\beta} 6 S_1 - e^{-6\beta} S_0\cr
f^F &=& (e^{-2\beta} - e^{-4\beta}) (S_0 + 3 S_2)
\eea
Unlensed indices are reproduced by taking $p = 1$ and are 
\bea
f^S &=& \frac{e^{-2\beta}+e^{-4\beta}}{(1-e^{-\beta})^5}\cr
f^T &=& \frac{10 e^{-5\beta} - 5 e^{-4\beta} + e^{-3\beta}}{(1-e^{-\beta})^5}\cr
f^F &=& 4\frac{e^{-2\beta}+e^{-3\beta}}{(1-e^{-\beta})^5}
\eea
and in total 
\bea
f^S+f^T-f^F &=& \frac{e^{-3\beta} - 3 e^{-2\beta}}{(1-e^{-\beta})^3}
\eea
Lensing gives
\bea
f^S+f^T-f^F &=& \frac{e^{-3p\beta}-(2+p^2)e^{-2p\beta}+(1-p^2)e^{-p\beta}}{(1-e^{-p\beta})^3}
\eea
but the expressions for the individual contributions are quite lengthy. Let us therefore only present their small $\beta$ expansions,
\bea
f^S &=& \frac{2}{p\beta^5} - \frac{1}{6 p \beta^3} + \frac{-160 + 168 p^2 + 21 p^4 + 2 p^6}{30240 p} \beta\cr
f^T &=& \frac{6}{p\beta^5} - \frac{5}{2 p \beta^3} + \frac{1}{p\beta} - \frac{1}{2} + \frac{-832 + 1848 p^2 - 63 p^4 + 2 p^6}{10080 p} \beta\cr
f^F &=& \frac{8}{p\beta^5} - \frac{2}{3p\beta^3} + \frac{-664 + 798 p^2 - 105 p^4 + 2 p^6}{7560 p}\beta
\eea
The leading order Stefan-Boltzmann terms are associated with
\bea
\zeta(6) &=& \frac{\pi^6}{945}
\eea
and are given by 
\bea
\beta F_S &=& -\frac{2}{945}\frac{\pi^6 r^5}{p\beta^5}\cr
\beta F_T &=& -\frac{2}{315}\frac{\pi^6 r^5}{p\beta^5}\cr
\beta F_F &=& -\frac{8}{945}\frac{\pi^6 r^5}{p\beta^5}
\eea
The Casimir energy contributions in the large $p$ limit are
\bea
E^S &=& -\frac{p^5}{30240r}\cr
E^T &=& -\frac{p^5}{10080r}\cr
E^F &=& -\frac{p^5}{7560r}
\eea
In the Stefan-Boltzmann behavior in the high temperature limit, we put $\beta = 2\pi r/k$, to get
\bea
\beta F^S &=& -\frac{2\pi k^5}{30240 p}\cr
\beta F^T &=& -\frac{2\pi k^5}{10080 p}\cr
\beta F^F &=& -\frac{2\pi k^5}{7560 p}
\eea
We would now like to reproduce this from the low temperature and the Casimir energy. We then take $\beta_D = 2\pi r/p$ and consider $L(k;1,1)$ and get
\bea
\beta_D E^S &=& -\frac{2\pi k^5}{30240 p}\cr
\beta_D E^T &=& -\frac{2\pi k^5}{10080 p}\cr
\beta_D E^F &=& -\frac{2\pi k^5}{7560 p}
\eea
We thus have got a complete agreement 
\bea
\beta F^i &=& \beta_D E^i
\eea
for all the individual fields in the tensor multiplet and so we have an asymptotic S-duality at order $T^5$ that appears to be generic. We also can see that at subleading orders, we have no such agreement, just as was to be expected.

\section{Discussion}
We have found that the unsquashing limit $a,b,c\rightarrow 0$ is smooth. In the 5d localization computation, the unsquashed case is difficult to analyse since the instanton particles will spread out as we take $a=b=c=0$ \cite{Kim:2012qf}. As long as the squashing parameters are not exactly zero, then no matter how small they are, once we fix their values and then take the localization limit, these instanton particles become localized at three fixed points on $\mb{C}P^2$,  in the localization limit. But we may worry that the unsquashed limit is singular, or that the limit is discontinuous and that we get a different result when we put $a=b=c=0$ compared to what we get by taking the limit $a,b,c\rightarrow 0$. Indeed such a discontinuity is natural to expect because we need a different computation when $a=b=c=0$ in the 5d theory. It is therefore a nontrivial result to have found that the limit $a,b,c\rightarrow 0$ is actually smooth. It would be interesting to see if this smooth behavior extends to nonabelian gauge groups.

\subsubsection*{Acknowledgements}
I would like to thank Seok Kim and Maxim Zabzine for discussions. This work was supported by the grant Geometry and Physics from Knut and Alice Wallenberg foundation.

\appendix
\section{The zeta function dualization method}
We follow \cite{Asorey:2012vp} and decompose the zeta function (\ref{zeta0}) into two parts,
\bea
\zeta(s) &=& \zeta_{\l=0} + 2 \zeta_{\l>0}(s)
\eea
where we notice that $\zeta_{\l<0} = \zeta_{\l>0}$, which explains the factor $2$. By a rewriting of the zeta function, we encounter coefficients 
\bea
C_{\lambda}(a_\l^2) &=& \sum_{\sigma} C_{\lambda,n} a_{\l}^{2\sigma}
\eea
that are polynomials in $a_{\l}^2$, where
\bea
a_{\l} &=& \frac{2\pi\l}{\beta}
\eea
Let us illustrate how such coefficient polynomials arise by a simple example. We consider the tensor multiplet zeta function (\ref{tzeta}) that (for $p=1$) involves terms of the form
\bea
(n^2-1) (n^2+a_{\l}^2)^{-s} &=& (n^2+a_{\l}^2-a_{\l}^2-1) (n^2+a_{\l}^2)^{-s}\cr
&=& (n^2+a_{\l}^2)^{1-s}-(a_{\l}^2+1) (n^2+a_{\l}^2)^{-s}
\eea
In this example, we have the coefficient polynomials
\bea
C_0(a_\l^2) &=& -1-a_{\l}^2\cr
C_1(a_\l^2) &=& 1
\eea
In a more general situtation, these coefficient polynomials are defined through the following expansions of the zeta function components above 
\bea
\zeta_{\l=0}(s) &=& (\mu r)^{2s} \sum_{\lambda} \sum_{n=1}^{\infty} C_{\lambda,0} (np)^{2\lambda-2s}\cr
&=& (\mu r)^{2s} \sum_{\lambda} C_{\lambda,0} p^{2\lambda-2s} \zeta(2s-2\lambda)
\eea
and
\bea
\zeta_{\l>0}(s) &=& (\mu r)^{2s} \sum_{\lambda} \sum_{\l=1}^{\infty} \sum_{n=1}^{\infty} C_{\lambda}(a_{\l}^2) \(a_{\l}^2+(np)^2\)^{\lambda-s}
\eea
respectively. We then apply the Mellin transform that puts $a_{\l}^2+(np)^2$ in the exponent,
\bea
\(a_{\l}^2+(np)^2\)^{\lambda-s} &=& \frac{1}{\Gamma(s-\lambda)} \int_0^{\infty} \frac{dt}{t} t^{s-\lambda} e^{-\(a_{\l}^2+n^2p^2\)t}
\eea
We now wish to dualize the sum with respect to $n$. To this end, we extend the sum over $n=1,2,...$ to also include $n=0,-1,-2,...$, which we can do since $n^2$ is even. But then we have to remember to subtract the term with $n=0$ again, and also divide the result by $2$. Once we have got a sum over $n\in \mb{Z}$, we can apply the Poisson resummation formula to that sum,
\bea
\sum_{n\in\mb{Z}} e^{-t n^2} &=& \sqrt{\frac{\pi}{t}} \sum_{n_D\in \mb{Z}} e^{-\frac{\pi^2 n_D^2}{t}}
\eea
For the terms with $n_D\neq 0$ we then use the following integral formula for the modified Bessel function
\bea
\int_0^{\infty} \frac{dt}{t} t^{\nu} e^{-\frac{a}{t}-bt} &=& 2\(\frac{a}{b}\)^{\nu/2} K_{\nu}(2\sqrt{ab})
\eea
while the integral we get for $n_D=0$ is gives a Gamma function. Then we must, as we have said, also subtract the term with $n=0$ (and then divide the whole thing by $2$). This way, we end up with the following result,
\bea
\zeta_{\l>0}(s) &=& \zeta_{\l>0,n=0}(s) + \zeta_{\l>0,n\neq 0}(s)
\eea
where
\bea
\zeta_{\l>0,n=0}(s) &=& \frac{\sqrt{\pi}}{2p}\mu^{2s} \sum_{\lambda,\sigma} C_{\lambda,\sigma} \frac{\Gamma(s-\lambda-1/2)}{\Gamma(s-\lambda)} \(\frac{2\pi}{\beta}\)^{1+2\lambda+2\sigma-2s} \zeta(2s-2\sigma-2\lambda-1)\cr
&&-\frac{1}{2}\mu^{2s} \sum_{\lambda,\sigma} C_{\lambda,\sigma} \(\frac{2\pi}{\beta}\)^{2\lambda+2n-2\sigma} \zeta(2s-2\sigma-2\lambda)
\eea
and
\bea
\zeta_{\l>0,n\neq 0}(s) &=& \frac{\sqrt{\pi}}{p} \mu^{2s} \sum_{\lambda} \sum_{\l=1}^{\infty} \sum_{n=1}^{\infty} \frac{C_{\lambda}}{\Gamma(s-\lambda)} \(\frac{n\beta}{2p\l}\)^{s-\lambda-1/2} K_{s-\lambda-1/2}\(\frac{4\pi^2 n\l}{p\beta}\)
\eea
Explicit forms of the Bessel functions that we will encounter are
\bea
K_{\pm 1/2}(x) &=& \sqrt{\frac{\pi}{2x}} e^{-x}\cr
K_{\pm 3/2}(x) &=& \sqrt{\frac{\pi}{2x}} \(1+\frac{1}{x}\) e^{-x}
\eea
To get the index, we need to compute the derivative of the zeta function at $s=0$. We have
\bea
\zeta'_{\l=0}(0) &=& \sum_{\lambda} C_{\lambda,0} p^{2\lambda} 2\zeta'(-2\lambda)
\eea
For $\zeta'_{\l>0}(0)$ we first bring up an overall factor of $s$ by using $1/\Gamma(s) = s/\Gamma(s+1)$ and $1/\Gamma(s-1) = s(s-1)/\Gamma(s+1)$. We then obtain the zeta function on the form $\zeta(s) = s \t \zeta(s)$ and the derivative is then simply given by $\zeta'(0) = \t \zeta(0)$ since in all our examples $\t \zeta(s)$ will be regular at $s=0$. Thus to compute the derivative, we never need to actually compute any derivative.

It would be more natural to apply Poisson resummation with respect to the sum over $\l$ that is already over $\mb{Z}$. But then we would get the low temperature expansion (\ref{lowT}) as was shown in \cite{Asorey:2012vp}. That computation is very elegant. In particular the Casimir energy factor drops out automatically without any need to consider normal ordering and zero point energies. 

\section{The Abel-Plana dualization method}
If we write the log of the index as a low temperature expansion
\bea
\log I &=& -\beta E + \sum_n d_n \log(1-e^{-\beta E_n})
\eea
then we may compute the sum over $n$ using the Abel-Plana integral formula. The analytic function that we need to consider in this application is given by
\bea
f(z) &=& d(z) \log(1-e^{-\beta E(z)})
\eea
where $d(n) = d_n$ and $E(n) = E_n$ for $n=0,1,2,...$, and analytically continued away from these integer values. If we compute the sum over $n$ by the Abel-Plana integral formula, then we automatically turn this sum into a high temperature expansion. 

Let us now present the Abel-Plana integration formula. A sum over $n$ may (under certain conditions) be computed by a contour integral,
\bea
\sum_{n=0}^{\infty} f(n) &=& \oint_C dz \frac{f(-z)}{e^{-2\pi i z} - 1}
\eea
Here $C$ is a counter clockwise contour surrounding the positive real axis, including the origin. Next we assume that $f$ is analytic in the positive halfplane and behave nicely at infinity, to rotate the contour to the imaginary axis. Then the right-hand side becomes
\bea
\frac{1}{2} f(0) + i\int_{-\infty}^{\infty} dx \frac{f(ix)}{e^{2\pi x} - 1}
\eea
where we add $f(0)/2$ because the integral contour that goes through the pole at $z=0$ picks up the other half of that same residue, to make up $f(0)/2+f(0)/2 = f(0)$ in total. We separate the integral domain into two pieces $[-\infty,\infty] = [-\infty,0]\cup[0,\infty]$ and rewrite the former integral
\bea
&& i\int_{-\infty}^0 dx \frac{f(ix)}{e^{2\pi x}-1}\cr
&=& i\int_0^{\infty} dx \frac{f(-ix)}{e^{-2\pi x}-1}\cr
&=& -i\int_0^{\infty} dx \frac{f(-ix)}{e^{2\pi x}-1}e^{2\pi x}\cr
&=& -i\int_0^{\infty} dx f(-ix) - i\int_0^{\infty} dx \frac{f(-ix)}{e^{2\pi x}-1}
\eea
The first integral is along the imaginary axis, but by assumption our function $f$ is well-behaved at infinity and aanalytic in the positive real halfplane, and the integral can be Wick rotated to the positive real axis,
\bea
-\int_0^{\infty} dx f(x) + i\int_0^{-\infty} dx f(ix) &=& 0
\eea
Adding up, we then have
\bea
\sum_{n=0}^{\infty} f(n) &=& \frac{1}{2} f(0) + \int_0^{\infty} dx f(x) + i\int_0^{\infty} dx \frac{f(ix)-f(-ix)}{e^{2\pi x}-1}
\eea
which is the Abel-Plana formula.

It is not apriori clear to us why the application of the Abel-Plana formula turns the low temperature expansion into a high temperature expansion, but it works this way in all explicit examples that we have encountered.

\section{The plethystic dualization method}
Here we describe the plethystic dualization method that was used in \cite{Kim:2012qf} to compute the index $I(\beta) = e^{-\beta E} \t I_{SB} \t I_{reg}$ where $E$ is the Casimir energy. We separate the generating function into a singular and a regular part at $\beta = 0$. The singular part goes into the Stefan-Boltzmann factor $\t I_{SB}$ as shown in the main text. The regular part is treated as follows,
\bea
\ln \t I_{reg}(\beta) = \sum_{n\in \mb{Z}} \int_{\epsilon}^{\infty} \frac{ds}{s} e^{\frac{2\pi i n s}{\beta}} f_{reg}(s)
\eea
We shall assume that $f_{reg}(s) = - f_{reg}(-s)$ and so $f_{reg}(s)/s$ does not have a simple pole at $s=0$. We put
\bea
\mu(n,s) &=& \frac{ds}{s} e^{\frac{2\pi i n s}{\beta}} f_{reg}(s)
\eea
where we notice that $\mu(-n,s) = \mu(n,-s)$. We then consider the following rewritings
\bea
\sum_{n\in \mb{Z}} \int_{\epsilon}^{\infty} \mu(n,ns) &=& \sum_{n=1}^{\infty} \int_{\epsilon}^{\infty} \mu(n,s)+ \sum_{n=1}^{\infty} \int_{\epsilon}^{\infty} \mu(-n,s) + \int_{\epsilon}^{\infty} \mu(0,s)\cr
&=& \sum_{n=1}^{\infty} \(\int_{-\infty}^{-\epsilon} \mu(n,s) + \int_{\epsilon}^{\infty} \mu(n,s)\) + \frac{1}{2} \(\int_{-\infty}^{-\epsilon} \mu(0,s) + \int_{\epsilon}^{\infty} \mu(0,s)\)\cr
&=& \sum_{n=1}^{\infty} \int_{\mb{R}} \mu(n,s) + \frac{1}{2} \int_{\mb{R}} \mu(0,s) -\sum_{n=1}^{\infty} \int_{-\epsilon}^{\epsilon} \mu(n,s) - \frac{1}{2} \int_{\epsilon}^{\epsilon} \mu(0,s)
\eea
We next look at the third and fourth terms,
\bea
-\sum_{n=1}^{\infty} \int_{-\epsilon}^{\epsilon} \mu(n,s)  - \frac{1}{2} \int_{-\epsilon}^{\epsilon} \mu(0,s) &=& -\frac{1}{2} \sum_{n\in\mb{Z}} \int_{-\epsilon}^{\epsilon} \mu(n,s)\cr
&=& - \frac{1}{2}\sum_{n\in\mb{Z}} \int_{-\epsilon}^{\epsilon} \frac{ds}{s} e^{\frac{2\pi i n s}{\beta}} f_{reg}(s)\cr
&=& -\frac{\beta}{2} \[\frac{1}{s} f_{reg}(s)\]\Bigg|_{s=0}
\eea
We thus have three terms to compute,
\bea
\ln \t I_{reg} &=& A + B + C
\eea
where
\bea
A &=& \sum_{n=1}^{\infty} \int_{\mb{R}} \frac{ds}{s} e^{\frac{2\pi i n s}{\beta}}f_{reg}(s)\cr
B &=& \frac{1}{2} \int_{\mb{R}} \frac{ds}{s} f_{reg}(s)\cr
C &=& -\frac{\beta}{2} \[\frac{1}{s} f_{reg}(s)\]\Bigg|_{s=0}
\eea
By noting that $f_{reg}(s) = - f_{reg}(-s)$ implies $f_{reg}(0) = 0$, we can write
\bea
C = -\frac{\beta}{2} \lim_{s\rightarrow 0} \[\frac{f_{reg}(s) - f_{reg}(0)}{s}\] = - \frac{\beta}{2} \lim_{s\rightarrow 0} \partial_s f_{reg}(s)
\eea
We now see that 
\bea
C &=& \beta E
\eea
where $E$ is the Casimir energy 
\bea
E &=& -\frac{1}{2} \partial_{\beta} f_{reg}(\beta)|_{\beta=0}
\eea
Then $e^C$ cancels against $e^{-\beta E}$. 

Let us try to illustrate this method by dualizing the Dedekind eta function. We start by rewriting its corresponding generating function as
\bea
f(\beta) = \frac{e^{-\beta}}{1-e^{-\beta}} = \frac{\cosh \frac{\beta}{2}}{2 \sinh \frac{\beta}{2}} - \frac{1}{2} 
\eea
Since the first term is antisymmetric, we can apply the plethystic dualization method on this term. The plethystic sum is given by 
\ben
\sum_{n=1}^{\infty} \frac{1}{n} f(n\beta) &=& \sum_{n=1}^{\infty} \(\frac{1}{2n} \frac{\cosh \frac{n\beta}{2}}{\sinh \frac{n\beta}{2}} - \frac{1}{2n}\)\label{PE}
\een
We note that although the whole expression is convergent, being equal to $-\sum_{n=1}^{\infty} \log\(1-e^{-n \beta}\)$, the sum $\sum_{n=1}^{\infty} -\frac{1}{2n}$ is divergent and has to be regularized if we shall be able to separate the two terms. But let us ignore this, and just apply the plethystic dualization method on the first term. From the singular piece, we get the Stefan-Boltzmann term
\bea
\sum_{n\in \mb{Z}} \int_{\epsilon}^{\infty} \frac{ds}{s} e^{2\pi i n s/\beta} \frac{1}{s} &=& \frac{\pi^2}{6\beta}
\eea
and from the regular piece we get the perturbative and nonperturbative contributions
\bea
&&\frac{1}{2}\int_{-\infty}^{\infty} \frac{ds}{s} \frac{\cosh \frac{s}{2}}{\sinh \frac{s}{2}} = \sum_{n=1}^{\infty} \frac{1}{2n}\cr
&&\int_{-\infty}^{\infty} \frac{ds}{s} \frac{\cosh \frac{s}{2}}{\sinh \frac{s}{2}} e^{-4\pi^2 k n/\beta} = \sum_{n=1}^{\infty} \frac{1}{n} e^{-4\pi^2 k n/\beta}
\eea
respectively. The perturbative contribution cancels against the second term in the plethystic sum (\ref{PE}), leaving us with  
\bea
\sum_{n=1}^{\infty} \frac{1}{n} f(n\beta) &=& \frac{\pi^2}{6\beta} + \sum_{k,n=1}^{\infty} \frac{1}{n} e^{-\frac{4\pi^2 k n}{\beta}}
\eea
The correct answer should have in addition a log-term $\frac{1}{2} \log \beta$, which we are missing. 

To make this computation rigorous, we may regularize the divergences by replacing $f(\beta)$ with $f(\beta) e^{-\epsilon \beta^2}$ and then at the end take $\epsilon \rightarrow 0$. We could alternatively use the zeta function dualization method and start with $\zeta(s) = \mu^{2s} \sum_{n\in \mb{N}} \sum_{\l \in \mb{Z}} \(a_{\l}^2+n^2\)^{-s}$ and obtain the high temperature expansion by dualizing with respect to $n$. We expect that the result will not depend on which regularization we use.

\end{document}